\documentclass[journal]{IEEEtran}

\usepackage{amssymb}
\usepackage[cmex10]{amsmath}
\usepackage{stfloats}
\usepackage{graphicx}
\usepackage{epstopdf}
\usepackage{subfigure}
\usepackage{tabularx}
\usepackage{epsfig,epsf,color,balance,cite}
\usepackage{bm}
\usepackage{multirow}
\usepackage{array}
\usepackage{setspace}
\usepackage{enumerate}
\usepackage{booktabs}
\usepackage{footnote}

\begin{document}
\begin{spacing}{1.0}

\title{Secrecy Outage and Diversity Analysis of Multiple Cooperative Source-Destination Pairs}
\author{Xiaojin Ding, Yulong Zou,~\IEEEmembership{Senior Member,~IEEE,} Xiaoshu Chen, Xiaojun Wang and Lajos Hanzo,~\IEEEmembership{Fellow,~IEEE}

\thanks{X. Ding is with the Jiangsu Engineering Research Center of Communication and Network Technology, Nanjing University of Posts and Telecommunications, Nanjing 210003, China. E-mail: dxj@njupt.edu.cn. X. Ding is also with the National Mobile Communications Research Laboratory, Southeast University, Nanjing 210096, China.}

\thanks{Y. Zou is with the Key Laboratory of Broadband Wireless Communication and Sensor Network Technology (Nanjing University of Posts and Telecommunications), Ministry of Education, Nanjing 210003, China. E-mail: yulong.zou@njupt.edu.cn}

\thanks{X. Chen, and X. Wang are with the National Mobile Communications Research Laboratory, Southeast University, Nanjing 210096, China. E-mail: dxj@njupt.edu.cn, \{xchen, wxj\}@seu.edu.cn}

\thanks{L. Hanzo is with the Department of Electronics and Computer Science, University of Southampton, Southampton, United Kingdom. (E-mail: lh@ecs.soton.ac.uk)}

}

\maketitle
\begin{abstract}
We study the physical-layer security of a multiple source-destination (SD) pairs coexisting wireless network in the face of an eavesdropper, where an eavesdropper intends to wiretap the signal transmitted by the SD pairs. In order to protect the wireless transmission against eavesdropping, we propose a cooperation framework relying on two stages. Specifically, an SD pair is selected to access the total allocated spectrum using an appropriately designed scheme at the beginning of the first stage. The other source nodes (SNs) simultaneously transmit their data to the SN of the above-mentioned SD pair relying on an orthogonal way during the first stage. Then, the SN of the chosen SD pair transmits the data packets containing its own messages and the other SNs' messages to its dedicated destination node (DN) in the second stage, which in turn will forward all the other DNs' data to the application center via the core network. We conceive a specific SD pair selection scheme, termed as the transmit antenna selection aided source-destination pair selection (TAS-SDPS). We derive the secrecy outage probability (SOP) expressions for the TAS-SDPS, as well as for the conventional round-robin source-destination pair selection (RSDPS) and non-cooperative (Non-coop) schemes for comparison purposes. Furthermore, we carry out the secrecy diversity gain analysis in the high main-to-eavesdropper ratio (MER) region, showing that the TAS-SDPS scheme is capable of achieving the maximum attainable secrecy diversity order. Additionally, increasing the number of the transmission pairs will reduce the SOP, whilst increasing the secrecy diversity order of the TAS-SDPS scheme. It is shown that the SOP of the TAS-SDPS scheme is better than that of RSDPS and Non-coop schemes. We also demonstrate that the secrecy diversity gain of proposed TAS-SDPS scheme is $M$ times that of the RSDPS scheme in the high-MER region, where $M$ is the number of the SD pairs.
\end{abstract}

\begin{IEEEkeywords}
Physical-layer security, source-destination pair scheduling, secrecy outage probability, secrecy diversity gain.
\end{IEEEkeywords}

\IEEEpeerreviewmaketitle

\section{Introduction}
\IEEEPARstart{M}{ultiple} source-destination pairs can be allowed to perform wireless transmissions simultaneously with the aid of spectrum sharing techniques [1]-[5], which are capable of increasing the system's efficiency and flexibility, whilst limiting interference imposed on each other. However, multiple source-destination (SD) pairs coexisting wireless systems may be vulnerable to both internal as well as to external attackers, when they operate independently in non-cooperative scenarios. For example, a hostile attacker may contaminate the legitimate transmission, thus degrading the quality of service (QoS). Furthermore, owing to the broadcast nature of radio propagation, the confidential messages may be overheard by malicious eavesdroppers. Hence, we have to protect wireless transmissions of the multiple SD coexisting systems against malicious eavesdropping.

Physical-layer security (PLS) [6]-[8] emerges as an effective method of guarding against wiretapping by exploiting the physical characteristics of wireless channels. Single-input multiple-output (SIMO) and multiple-input multiple-output (MIMO) schemes were conceived in [9], [10] for reducing the secrecy outage probability. Similarly, beamforming techniques were also invoked for improving the secrecy of wireless transmissions [11]-[12]. Moreover, the concept of cognitive jamming was explored in [13], while specially designed artificial noise was used for preventing eavesdropping in [14]. Furthermore, the authors of [15] and [16] explored opportunistic user scheduling conceived with cooperative jamming. More specifically, in [16], the non-scheduled users of the proposed user scheduling scheme were invoked for generating artificial noise in order to improve security in a multiuser wiretap network. Both one-way [17], [18] and two-way [19], [20] relaying schemes were conceived for guarding against eavesdropping, demonstrating that relay selection schemes are capable of improving the PLS. This is indeed expected, because they improve the quality of the desired link.

As a further development, PLS has also been designed for multi-system coexisting wireless networks, supporting a multiplicity of diverse devices. Hence, more efforts should be invested in enhancing the PLS of  wireless networks. The secrecy beamforming concept has been proposed by Lv et al. [21] for improving the PLS of heterogeneous networks. Moreover, jamming schemes have been investigated in [22]-[24]. To be specific, in [22], the jammers were selected to transmit jamming signals for contaminating the wiretapping reception of the eavesdroppers. Meanwhile, the interfering power imposed on the scheduled users was assumed to be below a threshold. A comprehensive performance analysis of artificial-noise aided secure multi-antenna transmission relying on a stochastic geometry framework was provided in [23] for $K$-tier heterogeneous cellar networks. In [24], joint beamforming and artificial noise scheme were designed at the secondary transmitters to guarantee secure wireless transmission. In [25], antenna selection was used for improving the security of source-destination transmissions in a multiple antenna aided MIMO system consisting of one source, one destination and one eavesdropper. In [26], a joint guard zone and threshold-based access control scheme was proposed for the D2D users to maximize the achievable secrecy throughput. Furthermore, the co-existence of a macro cell and a small cell constituting a simple cellular network was investigated by Zou [27]. Specifically, the overlay and underlay spectrum sharing schemes have been invoked for a macro cell and a small cell, respectively. Moreover, an interference-cancelation scheme was proposed for mitigating the interference in the underlay spectrum sharing case. In [28], Tolossa et al. investigated the base-station-user association scenarios suitable for protecting the ongoing transmission between the base-station and the intended user against eavesdropping. Additionally, the achievable average secrecy rate was analyzed by exploiting the association both with the ``best'' and with the $k$th best base-stations.

Against this backdrop, in this paper, we explore the PLS of a multiple SD pairs coexisting wireless network in the presence of an eavesdropper. In contrast to [21]-[28], we investigate the cooperation between different SD pairs for safeguarding against malicious eavesdropping with the aid of a specifically designed cooperative framework, and the main differences between this paper and [21]-[28] are summarized in table I. Moreover, we propose a pair of cooperation schemes based on source-destination (SD) pair scheduling. \textbf{\emph{More explicitly, against this background, the main contributions of this paper are summarized as follows.}} \begin{enumerate}

\item {Firstly, we propose a cooperative framework relying on two stages for protecting wireless transmissions against eavesdropping, Specifically, in the first stage, an SD pair will be chosen at the beginning of the transmission slot. Then, other source nodes (SNs) will confidentially transmit their data to the chosen SN via an orthogonal way. In the second stage, the specifically chosen SN transmits the repacked data to its destination node (DN), which will forward the received packets to the application center of the other SNs via the core network.}

\item {Secondly, we present a specific transmission selection scheme, termed as the transmit antenna selection aided source-destination pair scheduling (TAS-SDPS). To be specific, in the TAS-SDPS scheme, the ``best'' antenna of a chosen SD pair will be selected to transmit the repacked data relying on the total shared spectrum.}

\item {Thirdly, we analyze the secrecy outage probability (SOP) of the proposed TAS-SDPS scheme for transmission between SD pair over Rayleigh fading channels, whilst wireless transmission between SNs over Rician fading channels. We also evaluate the SOP of the traditional non-cooperative (Non-coop) and round-robin transmission pair scheduling (RSDPS) schemes for comparison. Moreover, we evaluate the secrecy diversity gains of both the TAS-SDPS and the RSDPS schemes, demonstrating that the TAS-SDPS scheme is capable of achieving the full secrecy diversity gain.}

\item {Finally, it is shown that the SOP of the TAS-SDPS scheme will be beneficially reduced by increasing the number of SD transmission pairs. Furthermore, the TAS-SDPS scheme outperforms the RSDPS and Non-coop schemes in terms of both the SOP and the secrecy diversity gain attained, demonstrating that the advantages of the proposed cooperative framework improves the security of wireless communications.} \end{enumerate}

The organization of this paper is as follows. In Section II, we briefly characterize the PLS of a multiple SD pairs coexisting wireless network. In Section III, we carry out the SOP analysis of the Non-coop, RSDPS, and TAS-SDPS schemes. In Section IV we evaluate the secrecy diversity gain of the proposed RSDPS and TAS-SDPS schemes. Our performance evaluations are detailed in Section VI. Finally, in Section V we conclude the paper.

\hfill

\hfill
\vspace{-2em}
\section{System Model and SD Pairs Scheduling}
\begin{table}
\tiny
\caption{Comparisons between our work and the related [21]-[28].}
\begin{tabular}{p{0.13\textwidth}|p{0.015\textwidth}|p{0.015\textwidth}|p{0.015\textwidth}|p{0.015\textwidth}|p{0.015\textwidth}|p{0.015\textwidth}|p{0.015\textwidth}|p{0.015\textwidth}|p{0.015\textwidth}}
\bottomrule
 & Our & [21]& [22]& [23]& [24]& [25]& [26]& [27]& [28]\\
\hline
Cooperative framework & \checkmark &  & & & & & & & \\
\hline
Spectrum-sharing & \checkmark & \checkmark &\checkmark & &\checkmark & & &\checkmark & \\
\hline
BS with multiple antennas & \checkmark & \checkmark & &\checkmark & &\checkmark & & &\checkmark \\
\hline
User with multiple antennas & \checkmark &  & & &\checkmark &\checkmark & & & \\
\hline
TAS-SDPS scheme & \checkmark &  & & & & & & & \\
\hline
Secrecy outage probability & \checkmark &  & & &\checkmark & & & & \\
\hline
Secrecy diversity gain & \checkmark &  & & & &\checkmark & &\checkmark & \\
\hline
Jamming &  &  &\checkmark &\checkmark &\checkmark & &\checkmark &\checkmark & \\
\hline
Secrecy rate & & \checkmark & & &\checkmark & &\checkmark & &\checkmark \\
\hline
Zero secrecy capacity & &  & & & &\checkmark & &\checkmark & \\
\hline
Connection probability and secrecy probability & &  &\checkmark &\checkmark & & &\checkmark & & \\
\hline
Against eavesdropping & \checkmark & \checkmark &\checkmark &\checkmark & &\checkmark & &\checkmark &\checkmark \\
\toprule
\end{tabular}
\end{table}

\begin{figure}
\centering
\includegraphics[width=0.85\linewidth]{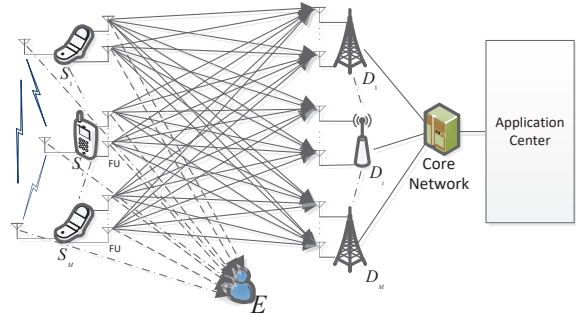}
\caption{A multiple SD pairs coexisting wireless network in the presence of an eavesdropper ${\text{E}}$.\label{fig:fig1}}
\end{figure}

\subsection{System Model}
As shown in Fig. 1, we consider $M$ source-destination (SD) pairs in the presence of an eavesdropper (E), where the E intends to wiretap the wireless transmissions of the legitimate source nodes (SNs) with the aid of a wide-band receiver. Each SN is assumed to be equipped with two radio frequency (RF) units, where one RF unit is used to perform wireless transmissions between the SNs, and the other one is invoked for communicating with the destination node (DN). For notational convenience, we let $\mathbb{D}$ represent the set of the SD pairs. Moreover, both the SNs-DNs and SNs-E links are modeled by Rayleigh fading [19], which are denoted by ${h_{{s_{{{m_i}}}}{d_{{{m_j}}}}}}$, ${h_{{s_{{{m_i}}}}{e_{l}}}}$ and ${h_{{s_{k}}{e_{l}}}}$, $m, k \in \{ {1, \cdots ,M} \}, k \ne m$, $i \in \{ {1, \cdots ,{N_T}} \}$, $j \in \{ {1, \cdots ,{N_R}} \}$, $l \in \{ {1, \cdots ,{N_E}} \}$, respectively, where ${N_T}$, ${N_R}$, and ${N_E}$ denote the number of antennas of the $S_m$ for communicating with the $D_m$, $D_m$, and $\text{E}$, respectively. The expected values of $|{h_{{s_{{{m_i}}}}{d_{{{m_j}}}}}}|^2$, $|{h_{{s_{{{m_i}}}}{e_{l}}}}|^2$ and $|{h_{{s_{k}}{e_{l}}}}|^2$ are $\sigma _{{s_{{m_i}}}{d_{{m_j}}}}^2$, $\sigma _{{s_{m_i}}{e_l}}^2$, and $\sigma _{{s_k}{e_l}}^2$, respectively. For notational convenience, upon denoting $\sigma _{{s_{{m_i}}}{d_{{m_j}}}}^2 = \alpha _{{s_{{m_i}}}{d_{{m_j}}}}^{}\sigma _{md}^2$, $\sigma _{{s_{m_i}}{e_l}}^2 = \alpha _{{s_{m_i}}{e_l}}^{}\sigma _{me}^2$, and $\sigma _{{s_k}{e_l}}^2 = \alpha _{{s_k}{e_l}}^{}\sigma _{me}^2$, where $\sigma _{md}^2$ and $\sigma _{me}^2$ are the respective reference channel gain of the SNs-DNs links and SNs-E links. Furthermore, we assume that all SNs are located in short range, and the links between SNs are characterized by Rician fading [29], which are represented by $({h_{{s_{{{k}}}}{s_{{{m}}}}}},{{K_{{s_k}{s_m}}}})$, where ${h_{{s_{{{k}}}}{s_{{{m}}}}}}$ and ${{K_{{s_k}{s_m}}}}$ are the instantaneous channel gain of $S_k\text{-}S_m$ link and the Rician K-factor of the $S_k\text{-}S_m$ link, $m, k \in \{ {1, \cdots ,M} \}, k \ne m$. Additionally, we assume that each SD pair can access an independent $B$Hz spectrum.

The cooperative framework relies on two stages. To be specific, at the beginning of the first stage, an SN will be chosen, which is chosen according to the criterion of the proposed TAS-SDPS scheme. Then, other SNs will simultaneously transmit their data to the appropriately selected SN using orthogonal resources (e.g., time-division, frequency-division, etc.). More specifically, if the chosen SN successfully decodes an SN's data, it will forward its data in the second stage. Otherwise, the SN's data will not be forwarded. Moreover, in the second stage, the SN chosen will concurrently retransmit its own data and along with the other SNs's data to its DN relying on orthogonal resources, where the DN will forward the received data to the application center through the core network.

\subsection{Signal Model}
In the first stage, let us assume that the SN $S_m$ is selected as the forwarding node. Then, other SNs will transmit their signals to $S_m$ on an orthogonal way with the aid of a single antenna, and $S_m$ receives all the rest SNs's data simultaneously. Without loss of generality, the signal received at $S_m$ transmitted by $S_k$, $k \in \mathbb{D} - \{ m \}$, is given by:
\begin{equation}
{y_{{s_{k}}{s_{m}}}} = \sqrt {{P_s}} {h_{{s_{k}}{s_{m}}}}{x_k} + {n_{{s_{m}}}}, \label{equ:1}
\end{equation}
where $P_s$, $x_k$, and ${n_{{s_{m}}}}$ denotes the transmitted power of $S_k$, the transmitted signal of $s_k$, and the thermal noise received at the $S_m$, respectively. In the meantime, the signal transmitted by $S_k$ will be overheard by $\text{E}$, which can be expressed as
\begin{equation}
{y_{{s_{k}}{e_l}}} = \sqrt {{P_s}} {h_{{s_{k}}{e_l}}}{x_k} + {n_{{e_l}}}, \label{equ:2}
\end{equation}
where ${n_{{e_l}}}$ represents the thermal noise received at $\text{E}$.

From (1) and (2), the achievable rate of the $S_k \text{-} S_m$ and $S_k \text{-} \text{E}$ links can be expressed as
\begin{equation}
{C_{{s_k}{s_m}}} = \frac{B}{2}{\log _2}( {1 + \gamma _{{s_k}{s_m}}} )\label{equ:3}
\end{equation}
and
\begin{equation}
{C_{{s_k}e}} = \frac{B}{2}{\log _2}\left( {1 + \gamma _{{s_k}e}^{}} \right),\label{equ:4}
\end{equation}
respectively, where $\gamma _{{s_k}{s_m}} ={\frac{P_s}{{{N_0}}}{{| {{h_{{s_{k}}{s_m}}}} |}^2}}$, $\gamma _{{s_k}e}^{} = \sum\limits_{l = 1}^{{N_E}} {\frac{{{P_s}{{| {{h_{{s_{k}}{e_l}}}} |}^2}}}{{{N_0}}}}$, $N_0$ denotes the variance of the thermal noise of $S_m$, $D_m$, and E, respectively.

In the second stage, $S_m$ transmits the successfully decoded data and its own data on an orthogonal way relying on $N_T$ antennas. Thus, following [33], the signal of an SN received at ${{D}_{m_j}}$ from ${{S}_{m}}$ can be formulated as
\begin{equation}
{y_{{s_{{m}}}{d_{{m_j}}}}} = \sum\limits_{i = 1}^{{N_T}}\sqrt {{P_{tx}}} {h_{{s_{{m_i}}}{d_{{m_j}}}}}{x_s} + {n_{{d_{{m_j}}}}}, \label{equ:5}
\end{equation}
where $P_{tx}$ and ${n_{{d_{{m}}}}}$ denote the transmitted power of the $S_{m_i}$, and the thermal noise received at the $D_{m_j}$, respectively. In the space-time coding (STC) case, for simplicity, we assume that the transmitted power of each antenna of $S_m$ is equal, thus, $P_{tx} = \frac{{{P_t}}}{N_T}$, where $P_t$ represents the available forwarding power of the RF unit of the $S_m$. By contrast, we have $P_{tx}=P_{t}$ in the transmit antenna selection (TAS) case. Without loss of generality, we assume that $E[|{x_k}|^2]=E[|{x_s}|^2]=1$, where $E[\cdot]$ denotes the operator of mathematical expectation. Similarly to (3), the signal transmitted by $S_m$ will be overheard by ${E_l}$, which can be written as
\begin{equation}
{y_{{s_{{m_i}}}{e_l}}} = \sum\limits_{i = 1}^{{N_T}}\sqrt {{P_{tx}}} {h_{{s_{{m_i}}}{e_l}}}{x_s} + {n_{{e_l}}}. \label{equ:6}
\end{equation}

Following [10] and [26], using (5) and maximal-ratio combining (MRC) [33], for each SN, the achievable rate of the $S_m \text{-} {D}_m$ and of the $S_{m_i} \text{-} {D}_m$ links in the STC and TAS cases can be formulated as
\begin{equation}
C_{{s_m}{d_m}}^{} = \frac{B}{2}{\log _2}\left( {1 + {\gamma _{{s_m}{d_m}}}} \right) \label{equ:7}
\end{equation}
and
\begin{equation}
C_{{s_{m_i}}{d_m}}^{} = \frac{B}{2}{\log _2}\left( {1 + {\gamma _{{s_{m_i}}{d_m}}}} \right), \label{equ:8}
\end{equation}
respectively, where ${\gamma _{{s_m}{d_m}}} = \sum\limits_{i = 1}^{{N_T}} {\sum\limits_{j = 1}^{{N_R}} {\frac{{P_t}{{| {{h_{{s_{{m_i}}}{d_{{m_j}}}}}} |}^2}}{{{N_T}{{N_0}}}} }}$, ${\gamma _{{s_{m_i}}{d_m}}} = {\sum\limits_{j = 1}^{{N_R}} {\frac{{P_t}{{| {{h_{{s_{{m_i}}}{d_{{m_j}}}}}} |}^2}}{{{{N_0}}}} }}$. It is pointed out that since the chosen SN retransmits concurrently its own data and the successfully decoded SNs's data relying on accessing the respective spectrum allocated to each SN, the actually effective achievable rate for each SD pair is given by (7) and (8), as $S_m$ transmits a packet consisting of all the data from the SNs.

Using (6) and MRC, for each SN, the achievable rate of the $S_m \text{-} \text{E}$ links can be expressed as
\begin{equation}
C_{{s_m}e} = \frac{B}{2}{\log _2}\left( {1 + \gamma _{{s_m}e}^{}} \right), \label{equ:9}
\end{equation}
where ${\gamma _{{s_m}{e}}} = \sum\limits_{i = 1}^{{N_T}} {\sum\limits_{l = 1}^{{N_E}} {\frac{{P_t}{{| {{h_{{s_{{m_i}}}{e_{l}}}}} |}^2}}{{{N_T}{{N_0}}}} }}$ in the STC case, and ${\gamma _{{s_{m}}{e}}} = {\sum\limits_{l = 1}^{{N_E}} {\frac{{P_t}{{| {{h_{{s_{{m_i}}}{e_{l}}}}} |}^2}}{{{{N_0}}}} }}$ in the TAS case.

Using (4) and (9), the overall capacity of the link spanning from $S_k$, $k \in \mathbb{D} - \{ m \}$, the wiretap channel from ${S_m} \text{-} {\text{E}}$ and ${S_k} \text{-} {\text{E}}$ can be obtained by using the maximum of the individual achievable rate of these two links in the first and second stages, i.e.
\begin{equation}
C_{{s}e}^{(k,m)} \!=\! \max \left( {C_{{s_k}e},C_{{s_m}e}} \right) \!=\! \frac{B}{2}{\log _2}\left[ {1 + \max \left( {\gamma _{{s_m}e}^{},\gamma _{{s_k}e}^{}} \right)} \right]. \label{equ:10}
\end{equation}

As mentioned above, given the chosen transmission pair, the signal of the chosen SN will only be transmitted during the second stage. By contrast, the signal of other SNs will be transmitted both during the first stage and be forwarded in the second stage. Hence, the signal of the other SNs that are being overheard in the two stages has been given in (3) and (5), respectively. Noting that although only selection combining (SC) is considered, here similar results can be achieved with the aid of MRC. Moreover, as discussed in [17], when independent and different codewords are used in the two stages, MRC becomes inapplicable, whereas SC is still suitable for the E.

\subsection{Transmit Antenna Selection Aided SD Pair Scheduling}
This subsection proposes a transmit antenna selection aided source-destination pair scheduling (TAS-SDPS) scheme. In the TAS-SDPS scheme, the ``best'' antenna having the maximal achievable rate of all SD pairs in the set $\mathbb{D}$ will be chosen to access the shared spectrum for the sake of improving the security of the SNs's wireless transmissions. Therefore, based on (7), the SD pair scheduling scheme in the TAS-SDPS can be formulated as
\begin{equation}
\left\{ {s,a} \right\}= \arg \mathop {\max }\limits_{m \in \mathbb{D}, 1 \le i \le {N_T}} C_{{s_{{m_i}}}{d_m}}, \label{equ:11}
\end{equation}
where $s$ represents the index of the selected pair in the TAS-SDPS scheme, and $a$ denotes the index of the chosen antenna of $S_s$, yielding:
\begin{equation}
\left\{ {s,a} \right\} = \arg \mathop {\max }\limits_{m \in \mathbb{D},1 \le i \le {N_T}} {{ {\sum\limits_{j = 1}^{{N_R}} {{{{{\left| {{h_{{s_{{m_i}}}{d_{{m_j}}}}}} \right|}^2}}}} } }}. \label{equ:12}
\end{equation}
Therefore, the secrecy capacity of $S_s \text{-} D_s$ and $S_k \text{-} D_s$ in the TAS-SDPS scheme can be formulated as $C_{\text{TAS}}^{s}={C_{{s_{s_a}}{d_s}}} - {C_{{s_s}e}}$ and $C_{{\rm{TAS}}}^k = \left\{ {\begin{array}{*{20}{c}}
{{C_{{s_{{s_a}}}{d_s}}} - C_{se}^{(k,m)}{\rm{ }}\;\;\text{if}\;\;{C_{{s_k}{s_s}}} > {R_o}}\\
\!\!\!\!\!\!\!\!\!\!\!\!\!\!{{C_{{s_k}{s_s}}} - C_{{s_k}e}^{}\;\;\;\;\;\text{otherwise}}
\end{array}} \right.$
, respectively.


\section{Secrecy outage probability analysis}
In this section, we present our performance analysis for the Non-coop, RSDPS, and TAS-SDPS schemes for transmission between SD pair over Rayleigh fading channels, whilst for transmission between SNs over Rician fading channels. The SOP expressions of the Non-coop scheduling as well as of the RSDPS and TAS-SDPS scheduling are derived.

\subsection{Conventional Non-coop Scheme}
For comparison, the traditional non-cooperative (Non-coop) transmission scheme is also presented, wherein each SN communicates with its DN independently. As above mentioned, each SN respectively occupies the $B$Hz channel bandwidth. Thus, different from (7) and (9), the instantaneous channel capacities of SN-DN and SN-E links are $B{\log _2}( {1 + {\gamma _{{s_m}{d_m}}}} )$ and $B{\log _2}( {1 + {\gamma _{{s_m}{e}}}} )$, respectively. The predefined secrecy rate of each SD pair is $R_s$. Hence, from (7) and (9), the SOP of the Non-coop scheme is expressed as
\begin{eqnarray}
&&\!\!\!\!\!\!\!\!\!\!\!\!\!\!\!\!\!\!\!\!\!\!P_{\text{so}}^{\text{Non}} \!\!=\!\! {\frac{1}{M}}\!\!\!\sum\limits_{m = 1}^{{M}}\!\!\!\Pr\!\! \left(\! B{\log _2}\!\left(\! {1 \!+\! {\gamma _{{s_m}{d_m}}}} \!\right) \!-\! B{\log _2}\!\left(\! {1 \!+\! {\gamma _{{s_m}{e}}}} \!\right) \!<\! {R_s} \!\right) \nonumber \\
&&\!\!\!\!\!\!\!\!\!\!\!\!\!\!\!\!\!\!\!\!\!\!=\!\!{\frac{1}{M}}\!\!\!\sum\limits_{m = 1}^{{M}}\!\!\Pr\!\! \left(\!\! {\sum\limits_{i = 1}^{{N_T}} {\sum\limits_{j = 1}^{{N_R}} \!\!{{{\left|\! {{h_{{s_{{m_i}}}{d_{{m_j}}}}}} \!\right|}^2}} } \!\! \!\!<\!\! {2^{{\frac{R_s}{B}}}}\!\sum\limits_{i = 1}^{{N_T}} \!{\sum\limits_{l = 1}^{{N_E}} \!{{{\left| {{h_{{s_{{m_i}}}{e_l}}}} \right|}^2}} }  \!\!+\!\! {\Delta _0^{'}}} \!\!\right),\label{equ:13}
\end{eqnarray}
where ${\Delta _0^{'}} = {{( {{2^{\frac{R_s}{B}}} - 1} ){N_T}{N_0}} \mathord{/
 {\vphantom {{( {{2^{\frac{R_s}{B}}} - 1} ){N_T}{N_0}} {{P_t}}}}
 \kern-\nulldelimiterspace} {{P_t}}}$, and according to (A.6), $P_{\text{so}}^{\text{Non}}$ can be obtained as
\begin{eqnarray}
&&\!\!\!\!\!\!\!\!\!\!\!\!\!\!\!\!\!\!\!\!\!P_{\text{so}}^{\text{Non}}\!\!=\!\!{\frac{1}{M}}\!\!\!\sum\limits_{m = 1}^{{M}}\!\!\!\left(\!\! 1 \!-\!\!\!\!\!\!\! \sum\limits_{l = 0}^{{N_T}{N_R} - 1} \!\!\!{\sum\limits_{p = 0}^l {\frac{{\left( {p + {N_T}{N_E} - 1} \right)!}}{{p!\left(\! {l - p} \!\right)!\left(\! {{N_T}{N_E} \!-\! 1} \!\right)!}}{{\left(\!\! {\frac{{{2^{{\frac{2 R_s}{B}}}}}}{{\sigma _{md}^2}}} \!\!\right)}^l}}}\right.\nonumber \\
&&\!\!\!\!\!\!\!\left.{\left(\!\! {\frac{1}{{\sigma _{me}^2}}} \!\!\right)^{{N_T}{N_E}}}\!\!\!\!{\left(\!\! {\frac{{{\Delta _0^{'}}}}{{{2^{{\frac{2 R_s}{B}}}}}}} \!\!\right)^{l - p}}\!\!{\left(\!\! {\frac{1}{{\sigma _{me}^2}} \!\!+\!\! \frac{{{2^{{\frac{2 R_s}{B}}}}}}{{\sigma _{md}^2}}} \!\!\right)^{ - p - {N_T}{N_E}}}\!\!\!\!\!\!\!\!\!\!{e^{ - \frac{{{\Delta _0^{'}}}}{{\sigma _{md}^2}}}}\right).\label{equ:14}
\end{eqnarray}

Observe from (13) and (14) that the conventional Non-coop scheme does not consider the cooperation between the SD pairs. Furthermore, it does not take the channel state information (CSI) of the SNs-DNs links into account. Although the Non-coop scheme is of lower complexity, it may degrade the PLS of the wireless transmission. Hence, this motivates us to conceive more advanced scheme for achieving SOP improvements.
\vspace{-1em}
\subsection{Conventional RSDPS Scheme}
This subsection provides the SOP analysis of the traditional RSDPS scheme used as a benchmarking scheme. In the conventional RSDPS scheme, each SD pair in the set $\mathbb{D}$ will be chosen to transmit with an equal probability. Therefore, according to the definition of SOP [8], we can obtain the SOP of the signal arriving from $S_m$ and $S_k$ in the first as well as second stage for the \text{RSDPS} scheme relying on the $S_m \text{-} D_m$ pair formulated as
\begin{equation}
P_{\text{so\_m\_m}}^{\text{RSDPS}} = \Pr \left( {C_{{s_m}{d_m}}} - {C_{{s_m}e}} < {R_s} \right) \label{equ:15}
\end{equation}
and
\begin{eqnarray}
&&\!\!\!\!\!\!\!\!\!\!\!\!\!\!\!\!\!\!\!\!\!\!\!\!\!P_{\text{so\_k\_m}}^{\text{RSDPS}} = \Pr \left( {{C_{{s_m}{d_m}}}-{C_{{s}e}^{(k,m)}}} <{R_s}, {C_{{s_k}{s_m}}}>R_o  \right) \nonumber \\
&&+\Pr \left({C_{{s_k}{s_m}}}-{C_{{s_k}e}}<R_s,{C_{{s_k}{s_m}}}<R_o\right), \label{equ:16}
\end{eqnarray}
respectively. Upon combining (7), (9) and (10), we arrive at
\begin{equation}
P_{\text{so\_m\_m}}^{\text{RSDPS}} \!\!=\!\! \Pr\!\! \left(\!\! {\sum\limits_{i = 1}^{{N_T}} {\sum\limits_{j = 1}^{{N_R}} {{{\left| {{h_{{s_{{m_i}}}{d_{{m_j}}}}}} \right|}^2}} } \!\!\! <\!\! {2^{{\frac{2 R_s}{B}}}}\!\!\sum\limits_{i = 1}^{{N_T}}\!\! {\sum\limits_{l = 1}^{{N_E}} \!\!{{{\left| {{h_{{s_{{m_i}}}{e_l}}}} \right|}^2}} }  \!\!+\!\! {\Delta _0}} \!\!\right) \label{equ:17}
\end{equation}
and
\begin{eqnarray}
&&\!\!\!\!\!\!\!\!\!\!\!\!\!\!\!P_{\text{so\_k\_m}}^{\text{RSDPS}} \!=\!\Pr\!\left(\!\!{{\left| {{h_{{s_{k}}{s_m}}}} \right|}^2}\!<\!{{2^{\frac{2\cdot R_s}{B}}}}\!\! \sum\limits_{l = 1}^{{N_E}}{{{{{\left| {{h_{{s_{k}}{e_l}}}} \right|}^2}}}}\!\!+\!\!\Theta_ 1, {{\left| {{h_{{s_{k}}{s_m}}}} \right|}^2}\!\!<\!\!\Theta_ 0  \!\!\right)\nonumber \\
&&\;\;+\!\Pr\!\! \left(\!\! {\sum\limits_{i = 1}^{{N_T}}\! {\sum\limits_{j = 1}^{{N_R}} {{{{{\left| {{h_{{s_{{m_i}}}{d_{{m_j}}}}}} \right|}^2}}}} }  \!\!\!<\!\! {\max}\!\left(\!\!{2^{{\frac{2 R_s}{B}}}}\sum\limits_{i = 1}^{{N_T}} {\sum\limits_{l = 1}^{{N_E}} {{{{{\left| {{h_{{s_{{m_i}}}{e_l}}}} \right|}^2}}}} },\right.}\right.\nonumber \\
&&\;\;\;\;\;\;\;\;\left.{\left.{\frac{{2^{{\frac{2 R_s}{B}}}}}{{{\Delta _1}}}}{\sum\limits_{l = 1}^{{N_E}} {{{{{\left| {{h_{{s_{k}}{e_l}}}} \right|}^2}}}} }\right) \!+\! {\Delta _0}} \!\right)\Pr\left(\!\!{{\left| {{h_{{s_{k}}{s_m}}}} \right|}^2}\!>\!\Theta_ 0  \!\!\right),\label{equ:18}
\end{eqnarray}
respectively, where we have ${\Delta _0} = {{( {{2^{\frac{2\cdot R_s}{B}}} - 1} ){N_T}{N_0}} \mathord{/
 {\vphantom {{( {{2^{\frac{2 \cdot R_s}{B}}} - 1} ){N_T}{N_0}} {{P_t}}}}
 \kern-\nulldelimiterspace} {{P_t}}}$, ${\Delta _1} = {{{P_t}} \mathord{/
 {\vphantom {{{P_t}} {( {{P_s}{N_T}} )}}}
 \kern-\nulldelimiterspace} {( {{P_s}{N_T}} )}}$, ${\Theta _0} = {{( {{2^{\frac{2\cdot R_o}{B}}} - 1} )} \mathord{/
 {\vphantom {{( {{2^{\frac{2 \cdot R_o}{B}}} - 1} )} {{\gamma_s}}}}
 \kern-\nulldelimiterspace} {{\gamma_s}}}$, ${\Theta _1} = {{( {{2^{\frac{2\cdot R_s}{B}}} - 1} )} \mathord{/
 {\vphantom {{( {{2^{\frac{2 \cdot R_s}{B}}} - 1} )} {{\gamma_s}}}}
 \kern-\nulldelimiterspace} {{\gamma_s}}}$, and $R_o$ is the data rate of a pair of SNs links. Furthermore, performing SD pair selection in the RSDPS scheme is independent of the random variables (RVs) ${| {{h_{{s_{{m_i}}}{d_{{m_j}}}}}}|^2}$ and ${| {{h_{{s_{{m_i}}}{e_l}}}}|^2}$. For simplicity, given the SD transmission pair $m$, we assume that the fading coefficients ${| {{h_{{s_{{m_i}}}{d_{{m_j}}}}}}|^2}$ for $i \in \{ {1,2, \cdots ,{N_T}} \}$, $j \in \{ {1,2, \cdots ,{N_R}} \}$, of all main channels are independent and identically distributed (i.i.d.) RVs with the same mean, denoted by $\sigma _{md}^2 = E( {{{| {{h_{{s_{{m_i}}}{d_{{m_j}}}}}} |}^2}} )$. Moreover, we also assume that the fading coefficients ${| {{h_{{s_{{m_i}}}{e_l}}}} |^2}$ for $i \in \{ {1,2, \cdots ,{N_T}} \}$, $l \in \{ {1,2, \cdots ,{N_E}} \}$ , of all wiretap links are i.i.d RVs having the same average channel gain denoted by $\sigma _{me}^2 = E( {{{| {{h_{{s_{{m_i}}}{e_l}}}} |}^2}} )$, which is a common assumption widely used in the cooperative communication literature. Hence, according to (A.6) and (A.10), (17) and (18) can be obtained as
\begin{eqnarray}
&&\!\!\!\!\!\!\!\!\!\!\!\!\!\!\!\!P_{\text{so\_m\_m}}^{\text{RSDPS}}\!\!=\!\!1 \!\!-\!\!\!\!\!\!\! \sum\limits_{l = 0}^{{N_T}{N_R} - 1} \!\!\!\!{\sum\limits_{p = 0}^l \!\!{\frac{{\left( {p + {N_T}{N_E} - 1} \right)!}}{{p!\left(\! {l \!-\! p} \!\right)!\left(\! {{N_T}{N_E} \!-\! 1} \!\right)!}}{{\left(\!\! {\frac{{{2^{{\frac{2 R_s}{B}}}}}}{{\sigma _{md}^2}}} \!\!\right)}^l}} }\!\!{\left(\!\! {\frac{1}{{\sigma _{me}^2}}} \!\!\right)^{{N_T}{N_E}}}\nonumber \\
&&\;\;\;\;\;\;{\left(\!\! {\frac{{{\Delta _0}}}{{{2^{{\frac{2 R_s}{B}}}}}}} \!\!\right)^{l - p}}{\left(\!\! {\frac{1}{{\sigma _{me}^2}} \!\!+\!\! \frac{{{2^{{\frac{2 R_s}{B}}}}}}{{\sigma _{md}^2}}} \!\!\right)^{ - p - {N_T}{N_E}}}\!\!\!\!\!{e^{ - \frac{{{\Delta _0}}}{{\sigma _{md}^2}}}}\label{equ:19}
\end{eqnarray}
and
\begin{eqnarray}
&&\!\!\!\!\!\!\!\!\!\!\!\!\!\!\!\!P_{\text{so\_k\_m}}^{\text{RSDPS}} = \!\!\bar P_{\text{o\_{km}}}\left(\!\!\sum\limits_{t = 0}^{{N_T}{N_E} \!-\! 1}\!\!\!\! {{\left(\!\!\frac{1}{\sigma _{ke}^2}\!\!\right)^{N_E}\!\!\!\!\left(\!\!\frac{1}{{\sigma _{me}^2}{\Delta _1}}\!\!\right)^{t} \!\!\frac{\left(\!t\!+\!N_E\!-\!1\!\right)!}{t!\left(\!N_E\!-\!1\!\right)!} } {{ {{c_{km}^{ - t\!-\! {N_E}}}} }}}\right.\nonumber \\
&&\left.-\!\!\!\!\!\! \sum\limits_{l = 0}^{{N_T}{N_R} - 1}\!\!\sum\limits_{p = 0}^{l}\!\!\sum\limits_{t = 0}^{p\!+\!{N_T}{N_E} \!-\! 1}\!\!\!\!\!\! {{a_{lp}{c_{md}}}{{ {{{\left(\!c_{km}\!+\!\frac{2^{\frac{2 R_s}{B}}}{{\Delta _1}{\sigma _{md}^2}}\!\right)}^{ - t- {N_E}}}} }}} \right. \nonumber \\
&& +\left.\!\!\!\sum\limits_{t = 0}^{{N_E} - 1}\!\!\! {{\left(\!\frac{1}{\sigma _{me}^2}\!\right)^{{N_T}{N_E}}\!\!\!\left(\!\frac{{\Delta _1}}{{\sigma _{ke}^2}}\!\right)^{t} \frac{\left(\!t\!+{\!N_T}{\!N_E}\!-\!1\!\right)!}{t!\left({\!N_T}{\!N_E\!}-\!1\!\right)!} } {{ {{d_{km}^{ - t\!-\! {N_T}{N_E}}}} }}}\right.\nonumber \\
&&\left. -\!\!\!\!\!\!\!\! \sum\limits_{l = 0}^{{N_T}{N_R} \!-\! 1}\!\!\!\!\sum\limits_{p = 0}^{l}\!\!\!\!\sum\limits_{t = 0}^{p\!+\!{N_E} \!-\! 1}\!\!\!\!\!\!\! {{a_{lp}{d_{kd}}}{{ {{{\left(\!\!d_{km}\!\!+\!\!\frac{2^{\frac{2 R_s}{B}}}{{\sigma _{md}^2}}\!\!\right)}^{ - t- {N_T}{N_E}}}} }}}\!\!\right)\!\!+\!\!P_{\text{{so}\_{km}}},\label{equ:20}
\end{eqnarray}
respectively, where $\bar P_{\text{o\_{km}}}$ and $P_{\text{{so}\_{km}}}$ are given by (A.8) and (A.9), respectively. Hence, the SOP of all SD pairs investigated relying on $S_m$ can be defined as
\begin{equation}
P_{\text{so\_m}}^{\text{RSDPS}} ={\frac{1}{M}}\left({\sum\limits_{k \in \mathbb{D} - \left\{ m \right\}}^{} {P_{\text{so\_k\_m}}^{\text{RSDPS}}}  + P_{\text{so\_m\_m}}^{\text{RSDPS}}}\right). \label{equ:21}
\end{equation}

As mentioned above, in the RSDPS scheme, each SD pair has an equal probability to be chosen. Furthermore, using the law of total probability [32], we can obtain the SOP for the RSDPS scheme as
\begin{equation}
P_{\text{so}}^{\text{RSDPS}} = \frac{1}{M}\sum\limits_{m = 1}^M {P_{\text{so\_m}}^{\text{RSDPS}}}. \label{equ:22}
\end{equation}

It is observed from (15) and (16) that although the RSDPS scheme considers the cooperation between the set of SNs, it is still independent of the CSIs of the SNs-DNs links, which implies that the employment of the TAS-SDPS scheme can further enhance the SOP of the wireless transmission in the wireless systems investigated.

\subsection{Proposed TAS-SDPS Scheme}
\vspace{-4em}
In this subsection, we present the SOP analysis of the TAS-SDPS scheme. As shown in (11), let $s$ denote the index of the chosen antenna of an SD pair under the TAS-SDPS scheme. Thus, we can formulate the SOP of the signal impinging from $S_s$ and $S_k$ under the TAS-SDPS scheme with the aid of the $S_s \text{-} D_s$ pair as
\begin{equation}
P_{\text{so\_s}}^{\text{TAS}} = \Pr \left( {{C_{{s_{s_a}}{d_s}}} - {C_{{s_{s_a}}e}} < {R_s} } \right) \label{equ:23}
\end{equation}
and
\begin{eqnarray}
&&\!\!\!\!\!\!\!\!\!\!\!\!\!\!\!\!\!P_{\text{so\_k}}^{\text{TAS}} = \Pr \left( {{C_{{s_{s_a}}{d_s}}} - {C_{{s}e}^{(k,s)}} < {R_s}}, {C_{{s_k}{s_s}}}>R_o \right)\nonumber \\
&&+\Pr \left({C_{{s_k}{s_s}}}-{C_{{s_k}e}}<R_s,{C_{{s_k}{s_s}}}<R_o\right), \label{equ:24}
\end{eqnarray}
respectively.

Using (8)-(10), both (23) and (24) can be rewritten as
\begin{equation}
P_{\text{so\_s}}^{\text{TAS}} = \Pr \left( {\sum\limits_{j = 1}^{{N_R}} {{{\left| {{h_{{s_{{s_a}}}{d_{s_j}}}}} \right|}^2}}  < {2^{\frac{2 R_s}{B}}}\sum\limits_{l = 1}^{{N_E}} {{{\left| {{h_{{s_{{s_a}}}{e_l}}}} \right|}^2}} +{\Lambda _0}} \right) \label{equ:25}
\end{equation}
and
\begin{eqnarray}
&&\!\!\!\!\!\!\!\!\!\!\!\!\!\!\!P_{\text{so\_k}}^{\text{TAS}} \!\!=\!\!\Pr\!\left(\!\!{{\left| {{h_{{s_{k}}{s_s}}}} \right|}^2}\!\!<\!\!{{2^{\frac{2\cdot R_s}{B}}}}\!\! \sum\limits_{l = 1}^{{N_E}}{{{{{\left| {{h_{{s_{k}}{e_l}}}} \right|}^2}}}}\!\!+\!\!\Theta_ 1, {{\left| {{h_{{s_{k}}{s_s}}}} \right|}^2}\!<\!\Theta_ 0  \right)\nonumber \\
&&\;\;+\!\Pr\! \left(\!\! { \sum\limits_{j = 1}^{{N_R}}\! {{{\left| {{h_{{s_{{s_a}}}{d_{s_j}}}}} \right|}^2}} \!\! <}{{2^{\frac{2 R_s}{B}}}\!\max\!\! \left(\!\! {\sum\limits_{l = 1}^{{N_E}} \!{{{\left| {{h_{{s_{{s_a}}}{e_l}}}} \right|}^2}} ,}\right.}\right.\nonumber \\
&&\;\;\;\;\;\;\;\;\;\;\left.{\left.{\frac{1}{{\Lambda _1}}\sum\limits_{l = 1}^{{N_E}} {{{\left| {{h_{{s_k}{e_l}}}} \right|}^2}} \!+\!{\Lambda _0}} \!\right)} \!\right)\Pr\left(\!\!{{\left| {{h_{{s_{k}}{s_s}}}} \right|}^2}\!>\!\Theta_ 0  \!\!\right), \label{equ:26}
\end{eqnarray}
respectively, where we have ${\Lambda _0} = {{( {{2^{\frac{2 \cdot R_s}{B}}} - 1} ){N_0}} \mathord{/
 {\vphantom {{( {{2^{\frac{2 \cdot R_s}{B}}} - 1} ){N_0}} {{P_t}}}}
 \kern-\nulldelimiterspace} {{P_t}}}$, and ${\Lambda _1} = {{{P_t}} \mathord{/
 {\vphantom {{{P_t}} { {{P_s}} }}}
 \kern-\nulldelimiterspace} { {{P_s}} }}$. Based on (12), we arrive at:
\begin{equation}
P_{\text{so\_s}}^{\text{TAS}} \!\!=\!\! \Pr\!\! \left(\!\! {\mathop {\max }\limits_{m \in \mathbb{D},1 \le i \le {N_T}} \sum\limits_{j = 1}^{{N_R}} {{{\left| {{h_{{s_{{m_i}}}{d_{m_j}}}}} \right|}^2}}  \!\!\!<\! }{{2^{\frac{2 R_s}{B}}}\!\sum\limits_{l = 1}^{{N_E}} {{{\left| {{h_{{s_{{m_i}}}{e_l}}}} \right|}^2}} \!\!+\!\!{\Lambda _0}} \!\!\right) \label{equ:28}
\end{equation}
and
\begin{eqnarray}
&&\!\!\!\!\!\!\!\!\!\!\!\!P_{\text{so\_k}}^{\text{TAS}} \!\!=\!\Pr\!\left(\!\!{{\left| {{h_{{s_{k}}{s_m}}}} \right|}^2}\!\!<\!\!{{2^{\frac{2\cdot R_s}{B}}}} \sum\limits_{l = 1}^{{N_E}}{{{{{\left| {{h_{{s_{k}}{e_l}}}} \right|}^2}}}}\!+\!\Theta_ 1, {{\left| {{h_{{s_{k}}{s_m}}}} \right|}^2}\!\!<\!\!\Theta_ 0  \!\!\right)\nonumber \\
&&+\!\!\Pr\!\! \left(\!\! {\mathop {\max }\limits_{m \in \mathbb{D},1 \le i \le {N_T}} \!\!\!\sum\limits_{j = 1}^{{N_R}}\! {{{\left| {{h_{{s_{{m_i}}}{d_{m_j}}}}} \right|}^2}}  \!\!\!\!\!<\!\!{2^{\frac{2 R_s}{B}}}\!\!\max\!\! \left(\!\! {\sum\limits_{l = 1}^{{N_E}} \!{{{\left| {{h_{{s_{{m_i}}}{e_l}}}} \right|}^2}} ,}\right.}\right.\nonumber \\
&&\;\;\;\;\;\;\;\left.{\left.{\frac{1}{\Lambda _1}\!\sum\limits_{l = 1}^{{N_E}} {{{\left| {{h_{{s_k}{e_l}}}} \right|}^2}} \!+\!{\Lambda _0}} \!\right)} \!\right)\Pr\left(\!\!{{\left| {{h_{{s_{k}}{s_s}}}} \right|}^2}\!>\!\Theta_ 0  \!\!\right), \label{equ:29}
\end{eqnarray}
respectively.

Finally, using (A.20) and (A.21), both (27) and (28) can be obtained as
\begin{equation}
P_{\text{so\_s}}^{\text{TAS}}= \sum\limits_{S'} {\sum\limits_{p = 0}^{{\beta _2}} {{\Psi _0}\left( {p + {N_E} - 1} \right)!{{\left( {\frac{1}{{\sigma _{me}^2}} + {\beta _3}{2^{{\frac{2 R_s}{B}}}}} \right)}^{ - p - {N_E}}}} } \label{equ:29}
\end{equation}
and
\begin{eqnarray}
&&\!\!\!\!\!\!\!\!\!\!\!\!\!\!\!\!\!\!\!\!\!P_{\text{so\_k}}^{\text{TAS}} \!\!=\! \bar P_{\text{o\_{km}}}\!\left(\!\! {\sum\limits_S {\sum\limits_{p = 0}^{\beta _2}\!\! \sum\limits_{t = 0}^{p \!+\! {N_E} \!-\! 1}\!\!\!\!\!{a_{{\beta}{p}}{c_{{\beta}d}}{{\left(\!\! \frac{d_{km}^{'}}{\Lambda _1} \!+\! \frac{{\beta _3}{2^{\frac{2 R_s}{B}}}}{\Lambda _1} \!\!\right)}^{ - t \!-\! {N_E}}}} } } \right. \nonumber \\
&&\!\!\!\!\!\!\!\!\!\!\!\left.+\!  {\sum\limits_S \!{\sum\limits_{p = 0}^{\beta _2}\!\!\sum\limits_{t = 0}^{p\!+\! {N_E} \!-\! 1}\!\!\!\!\! {a_{{\beta}{p}}{d_{{\beta}d}}{{\left(\!\! {c_{km}^{'}}{\Lambda _1} \!\!+\!\! {{\beta _3}{2^{\frac{2 R_s}{B}}}} \!\!\right)}^{ - t \!-\! {N_E}}}} } }\!\!\right)\!\!+\!\!P_{\text{{so}\_{km}}}, \label{equ:30}
\end{eqnarray}
respectively. Moreover, relying on the definition in (21), the SOP of the investigated system relying on the proposed TAS-SDPS scheme can be expressed as:
\begin{equation}
P_{\text{so}}^{\text{TAS}} ={\frac{1}{M}}\left({\sum\limits_{k \in \mathbb{D} - \left\{ s \right\}}^{} {P_{\text{so\_k}}^{\text{TAS}}}  + P_{\text{so\_s}}^{\text{TAS}}}\right) . \label{equ:31}
\end{equation}

So far, we have derived closed-form SOP expressions of the conventional Non-coop and RSDPS schemes as well as the proposed TAS-SDPS scheme.

\section{Secrecy Diversity Gain Analysis}
In this section, we present the secrecy diversity analysis of the RSDPS and TAS-SDPS schemes in the high MER region for the sake of providing further insights from (17), (18), (25) and (26) conceiving both the conventional RSDPS as well as the proposed TAS-SDPS scheme.

\subsection{Traditional RSDPS Scheme}
This subsection analyzes the asymptotic SOP of the conventional RSDPS scheme. In the spirit of [27], the traditional diversity gain is defined in [31] as
\begin{equation}
d =  - \mathop {\lim }\limits_{\text{SNR} \to \infty } \frac{{\log {P_\text{e}}\left( \text{SNR} \right)}}{{\log \text{SNR}}}, \label{equ:32}
\end{equation}
which is used for characterizing the reliability of wireless communications, where SNR and ${P_e}( \text{SNR})$ denote the signal-to-noise ratio (SNR) of the destination node and the bit error ratio (BER), respectively. However, we can observe that the SOPs of the RSDPS and TAS-SDPS schemes are independent of the SNR, hence the definition of the traditional diversity gain may not perfectly suit our SOP analysis. Moreover, as shown in (17), (18), (25) and (26), the SOP of the RSDPS scheme is related to the main channel ${|\! {{h_{{s_{{m_i}}}{d_{{m_j}}}}}} \!|^2}$ as well as to the eavesdropping channels ${|\! {{h_{{s_{{m_i}}}{e_l}}}} \!|^2}$ and ${| {{h_{{s_{k}}{e_l}}}} |^2}$. For notational convenience, let ${\lambda _{se}} \!=\! {{\sigma _{md}^2} \mathord{/
 {\vphantom {{\sigma _{md}^2} {\sigma _{me}^2}}}
 \kern-\nulldelimiterspace} {\sigma _{me}^2}}$ denote MER. In spirit of the above observation, and following [8] and [25], we define the secrecy diversity gain as the asymptotic ratio of the logarithmic SOP to the logarithmic ${\lambda _{se}}$ as ${\lambda _{se}} \!\to\! \infty $, which is mathematically formulated as
\begin{equation}
d =  - \mathop {\lim }\limits_{{\lambda _{se}} \to \infty } \frac{{\log \left( {{P_{\text{so}}}} \right)}}{{\log \left( {{\lambda _{se}}} \right)}}. \label{equ:33}
\end{equation}

Meanwhile, in (33), the SOP ${P_{\text{so}}}$ behaves as $\lambda _{se}^{ - d}$ in the high MER region, which means that upon increasing the diversity gain $d$, ${P_{\text{so}}}$ decreases faster in the high MER region. Using (33), the secrecy diversity gain of the RSDPS scheme can be expressed as
\begin{equation}
{d_{{\text{RSDPS}}}} =  - \mathop {\lim }\limits_{{\lambda _{se}} \to \infty } \frac{{\log \left( { P_{\text{so}}^{\text{RSDPS}}} \right)}}{{\log \left( {{\lambda _{se}}} \right)}}. \label{equ:34}
\end{equation}

\emph{Theorem 1:} The secrecy diversity gain of the RSDPS scheme is given by
\begin{equation}
{d_{{\text{RSDPS}}}} = {N_T}{N_R}. \label{equ:35}
\end{equation}
\;\;\;\;\;\;\;\;\emph{Proof:} Please refer to Appendix B.

\emph{Remark 1:} We can observe from Theorem 1 that the RSDPS scheme only attains a secrecy diversity gain of ${N_T}{N_R}$, and the SOP of the RSDPS scheme is governed by the factor ${( {{\textstyle{1 \over {{\lambda _{se}}}}}} )^{{N_T}{N_R}}}$ in the high-MER region. This is due to the fact that the secrecy diversity gain of the RSDPS scheme only depends on the number of antennas involved by a pair of the transmitters and receivers. Since ${d_{{\text{RSDPS}}}}$ does not depend on the number of SD pairs, the RSDPS scheme achieves no SOP enhancement upon increasing the number of SD pairs, which is a disadvantage of the RSDPS scheme. Moreover, the secrecy diversity gain of the Non-coop scheme can be similarly obtained as ${N_T}{N_R}$.

\subsection{Proposed TAS-SDPS Scheme}
This subsection is focused on the secrecy diversity analysis of the TAS-SDPS scheme. Similarly to (34), the secrecy diversity order of the TAS-SDPS scheme can be expressed as
\begin{equation}
{d_{\text{TAS}}} =  - \mathop {\lim }\limits_{{\lambda _{se}} \to \infty } \frac{{\log \left( { P_{\text{so}}^{\text{TAS}}} \right)}}{{\log \left( {{\lambda _{se}}} \right)}}. \label{equ:36}
\end{equation}

\emph{Theorem 2:} The secrecy diversity gain of the TAS-SDPS scheme yields to
\begin{equation}
{d_{{\text{TAS}}}} = M{N_T}{N_R}. \label{equ:49}
\end{equation}
\;\;\;\;\;\;\;\;\emph{Proof:} Please refer to Appendix B.

\emph{Remark 2:} Interestingly, we can see from Theorem 2 that the TAS-SDPS scheme achieves the secrecy diversity gain of $M{N_T}{N_R}$, which means that the SOP of the TAS-SDPS scheme is governed by the factor ${( {{\textstyle{1 \over {{\lambda _{se}}}}}} )^{M{N_T}{N_R}}}$ in the high-MER region. The SOP of the TAS-SDPS scheme can be improved not only by increasing the number of antennas of a transmitter and receiver pair, but also by increasing the number of the SD pairs. Therefore, the TAS-SDPS scheme advocated significantly outperforms the conventional RSDPS and Non-coop scheme in terms of their SOPs.
\section{Performance evaluation}
In this section, we present our performance comparisons among the Non-coop, the RSDPS, the proposed TAS-SDPS schemes in terms of their SOPs and secrecy diversity gains. Specifically, the analytic SOPs of the Non-coop, the RSDPS, and TAS-SDPS schemes are evaluated by plotting (14), (22) and (31), respectively. Moreover, the lower bound SOPs of the RSDPS and TAS-SDPS schemes are obtained by using (B.15), and (B.24), respectively. The upper bound SOP of the RSDPS and TAS-SDPS schemes are obtained by using (B.18), and (B.27), respectively. The simulated SOP of the RSDPS as well as the proposed the TAS-SDPS schemes are also provided for demonstrating the correctness of the theoretical results. In our numerical evaluation, we assume that $\alpha _{{s_{{m_i}}}{e_l}}^{}$ = $\alpha _{{s_{{k}}}{e_l}}^{}$ = $\alpha _{{s_{{m_i}}}{d_{{m_j}}}}^{}$ = 1.

\begin{figure}
\centering
\includegraphics[width=0.9\linewidth]{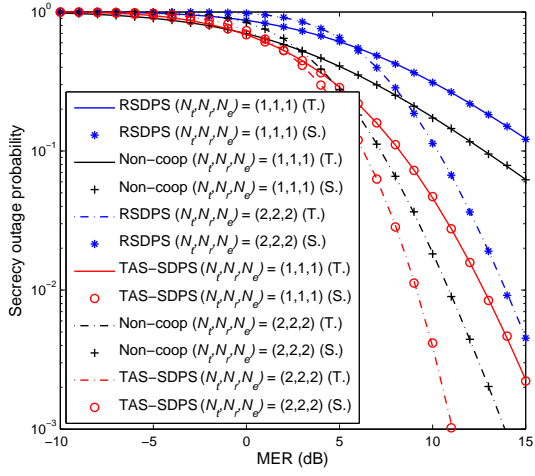}
\caption{SOP vs MER ${\lambda _{se}}$ of the traditional Non-coop and RSDPS as well as the proposed TAS-SDPS schemes for different $(N_T,N_R,N_E)$ with $M$ = 4.\label{fig:fig2}}
\end{figure}

\begin{figure}
\centering
\includegraphics[width=0.9\linewidth]{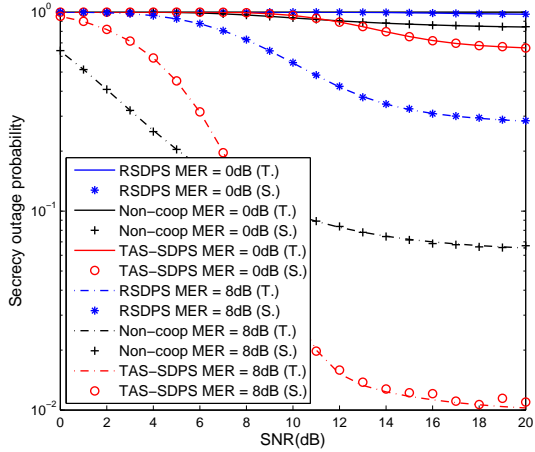}
\caption{SOP vs SNR ${{\textstyle{{{P_t}} \over {{N_0}}}}}$ of the traditional Non-coop and RSDPS as well as the proposed TAS-SDPS schemes for different $MER$ with $N_T$ = $N_R$ = $N_E$ = 2, and $M$ = 8.\label{fig:fig3}}
\end{figure}

\begin{figure}
\centering
\includegraphics[width=0.94\linewidth]{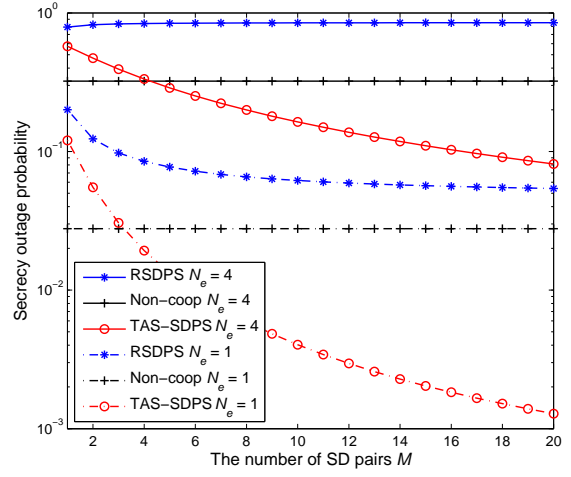}
\caption{SOP vs the number of source-destination pairs $M$ of the traditional Non-coop and RSDPS as well as the proposed TAS-SDPS schemes for different $N_E$ with $N_T$ = $N_R$ = 2, and ${\lambda _{se}}$ = 10dB.\label{fig:fig4}}
\end{figure}

\begin{figure}
\centering
\includegraphics[width=0.89\linewidth]{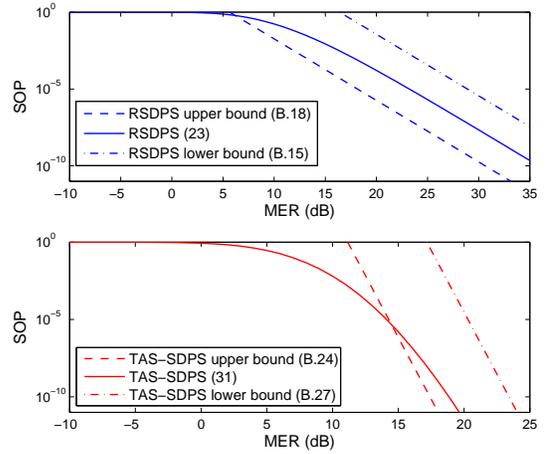}
\caption{Asymptotic and exact results on the SOP of the traditional RSDPS as well as the proposed TAS-SDPS schemes with $N_T$ = $N_R$ = $N_E$ = 2, and $M$ = 4.\label{fig:fig5}}
\end{figure}
In Fig. 2, we show the SOP versus MER ${\lambda _{se}}$ of both the traditional Non-coop and of the RSDPS as well as of the proposed TAS-SDPS schemes for different parameters $(N_T,N_R,N_E)$ by plotting (15), (22) and (31), as a function of the MER ${\lambda _{se}}$. It is shown in Fig. 2 that the SOPs of the RSDPS, of the Non-coop, and of the TAS-SDPS schemes decrease, as the number of antennas $(N_T,N_R,N_E)$ increases from $(N_T,N_R,N_E) = (1,1,1)$ to $(2,2,2)$. Furthermore, the RSDPS, the Non-coop, and the TAS-SDPS schemes using $(N_T,N_R,N_E) = (2,2,2)$ achieve better secrecy performance than that of $(N_T,N_R,N_E) = (1,1,1)$, respectively. Fig. 2 also demonstrates that increasing the MER upgrades the security of wireless transmissions in networks. Additionally, Fig. 2 demonstrates that the TAS-SDPS scheme attains the best SOP performance among the traditional RSDPS and Non-coop as well as the proposed TAS-SDPS schemes, when the MER increases from -10dB to 15dB.

Fig. 3 illustrates the SOP versus the SNR ${{\textstyle{{{P_t}} \over {{N_0}}}}}$ of the traditional RSDPS and of Non-coop as well as of the proposed TAS-SDPS schemes. Fig. 3 shows that increasing the SNR ${{\textstyle{{{P_t}} \over {{N_0}}}}}$ may moderately degrade the SOPs of the RSDPS, of the Non-coop as well as of the proposed TAS-SDPS schemes in the MER = 0dB case. By contrast, upon increasing the SNR, the SOPs of all schemes decreases are significantly reduced in the MER = 8dB case. This can be explained by observing that increasing the SNR is beneficial both for the SNs-DNs links and for the SNs-E links in the MER = 0dB case. However, increasing the SNR may be more beneficial for the SNs-DNs links than for the SNs-E links in the MER = 8dB case. Furthermore, it can also be seen from Fig. 3 that the SOP of the proposed TAS-SDPS scheme is lower than that of the RSDPS and Non-coop schemes at a specific SNR. In contrast to the Non-coop and RSDPS schemes, this means that the security performance benefits from exploiting the cooperation between the SD pairs by guarding against eavesdropping with the aid of proposed TAS-SDPS scheme.

Fig. 4 shows our SOP comparison of the traditional RSDPS and Non-coop as well as of the proposed TAS-SDPS schemes for different number of the SD pairs $M$. Observe from Fig. 4 that as the number of SD pairs increases from $M$ = 2 to 20, the SOP of the TAS-SDPS scheme is reduced significantly, which shows that increasing the number of SD pairs is beneficial for the PLS of the proposed TAS-SDPS scheme, both in the cases of $N_e$ = 1 and $N_e$ = 4. This is due to the fact that when $M$ increases from $M$ = 2 to 20, the proposed TAS-SDPS scheme can take advantage of the cooperation between different SD pairs for enhancing the PLS of wireless networks. However, the SOPs of the RSDPS and of the Non-coop schemes remain unchanged, when the number of SD pairs increases from $M$ = 2 to 20. Moreover, upon an increasing $N_e$, the SOP of the TAS-SDPS scheme can be updated by increasing the number SD pairs $M$. As shown in Fig. 4, the proposed TAS-SDPS scheme outperforms the Non-coop and RSDPS schemes in terms of their SOPs for all the $M$ values.

Fig. 5 shows both the asymptotic and the exact results conceiving the SOP of the traditional RSDPS as well as of the proposed TAS-SDPS schemes, where the lower bound results, exact results and the upper bound results are obtained by plotting (B.15), (B.24), (22), (31), (B.18), and (B.27) as a function of the MER, respectively. Observe from Fig. 5 that the exact SOP curves of the RSDPS, and the TAS-SDPS schemes are more and more close to their corresponding lower and upper bounds, as the MER increases. Moreover, as shown in Fig. 5, in the high-MER region, the exact SOP curves of the RSDPS, and TAS-SDPS schemes exhibit the same slopes of their corresponding lower and upper bounds, respectively. This demonstrates the correctness of our secrecy diversity gain analysis of the RSDPS, and TAS-SDPS schemes in the high-MER region.
\section{Conclusions}
In this paper, we explored a wireless network coexisting with multiple wireless systems in the face of an eavesdropper, supporting multiple SD pairs, where each SD pair may access the shared spectrum dynamically, and the eavesdropper aims for maliciously wiretapping the signals transmitted by the user nodes relying on a wide-band receiver. We proposed a cooperative framework relying on two stages for enhancing the PLS of the ongoing wireless transmissions, wherein an SD pair will be chosen as the transmitting pair within a given spectral band from the perspective of security. Moreover, we presented an SD pair scheduling scheme, which is termed as the TAS-SDPS. We analyzed the SOP of the proposed TAS-SDPS scheme, and carried out the SOP analysis of both the RSDPS and of the Non-coop schemes as a baseline. We also carried out the secrecy diversity gain analysis of the TAS-SDPS scheme, as well as of the RSDPS scheme. It was demonstrated that the TAS-SDPS scheme outperforms both the RSDPS and the Non-coop schemes in terms of their SOPs. Furthermore, as the number of SD pairs increases, the SOP of the TAS-SDPS scheme improves, while the SOPs of the RSDPS and Non-coop schemes remain unchanged.

\numberwithin{equation}{section}
\appendices
\section{}
Upon defining $U={\sum\limits_{i = 1}^{{N_T}} {\sum\limits_{j = 1}^{{N_R}} {{{{{| {{h_{{s_{{m_i}}}{d_{{m_j}}}}}} |}^2}}}} }}$, $X_1 = \sum\limits_{i = 1}^{{N_T}}\! {\sum\limits_{l = 1}^{{N_E}}\! {{{{{| {{h_{{s_{{m_i}}}{e_l}}}} |}^2}}}} }$, $X_2={\sum\limits_{l = 1}^{{N_E}} {{{{{| {{h_{{s_{k}}{e_l}}}} |}^2}}}} }$, and $X_3={ {{{{{| {{h_{{s_{k}}{s_m}}}} |}^2}}}} }$, and taking into account that the RVs ${{| {{h_{{s_{{m_i}}}{d_{{m_j}}}}}} |}^2}$, ${{| {{h_{{s_{k}}{e_l}}}} |}^2}$, ${{| {{h_{{s_{{m_i}}}{e_l}}}} |}^2}$, and ${ {{{{{| {{h_{{s_{k}}{s_m}}}} |}^2}}}} }$ are independent of each other, $P_{\text{so\_m\_m}}^{\text{RSDPS}}$ and $P_{\text{so\_k\_m}}^{\text{RSDPS}}$ can be expressed as
\begin{eqnarray}
&&\!\!\!\!\!\!\!\!\!\!\!\!\!P_{\text{so\_m\_m}}^{\text{RSDPS}} \!=\! \Pr\! \left(\! {\sum\limits_{i = 1}^{{N_T}} {\sum\limits_{j = 1}^{{N_R}} {{{\left| {{h_{{s_{{m_i}}}{d_{{m_j}}}}}} \right|}^2}} }  \!<\! {2^{{\frac{2 R_s}{B}}}}\sum\limits_{i = 1}^{{N_T}} {\sum\limits_{l = 1}^{{N_E}} {{{\left| {{h_{{s_{{m_i}}}{e_l}}}} \right|}^2}} }  \!\!+\!\! {\Delta _0}} \!\right) \nonumber \\
&&\;\;\;\;= \int_0^\infty  { {{F_U}\left( {\Delta _0}+2^{\frac{2 R_s}{B}}{x_1} \right)} } {f_{X_1}}\left( {x_1} \right)d{x_1}
\end{eqnarray}
and
\begin{eqnarray}
&&\!\!\!\!\!\!\!\!\!\!\!\!\!\!\!P_{\text{so\_k\_m}}^{\text{RSDPS}} \!=\! \Pr\!\! \left(\!\! {\sum\limits_{i = 1}^{{N_T}} \!{\sum\limits_{j = 1}^{{N_R}}\! {{{{{\left|\! {{h_{{s_{{m_i}}}{d_{{m_j}}}}}} \!\right|}^2}}}} }  \!\!\!\!<\! {\max}\!\left(\!\!{2^{{\frac{2 R_s}{B}}}}\sum\limits_{i = 1}^{{N_T}} {\sum\limits_{l = 1}^{{N_E}} {{{{{\left| {{h_{{s_{{m_i}}}{e_l}}}} \right|}^2}}}} }\right.}\right.,\nonumber \\
&&\;\;\;\;\;\;\;\;\;\;\;\left.{\left.{\frac{{2^{{\frac{2 R_s}{B}}}}}{{{\Delta _1}}}}{\sum\limits_{l = 1}^{{N_E}} {{{{{\left| {{h_{{s_{k}}{e_l}}}} \right|}^2}}}} }\right) \!+\! {\Delta _0}} \!\!\right)\Pr\left(\!\!{{\left| {{h_{{s_{k}}{s_m}}}} \right|}^2}\!>\!\Theta_ 0  \!\!\right) \nonumber \\
&&\;\;+\Pr\!\!\left(\!\!{{\left|\! {{h_{{s_{k}}{s_m}}}} \!\right|}^2}\!\!\!<\!{{2^{\frac{2\cdot R_s}{B}}}} \sum\limits_{l = 1}^{{N_E}}{{{{{\left| {{h_{{s_{k}}{e_l}}}} \right|}^2}}}}\!+\!\Theta_ 1, {{\left| {{h_{{s_{k}}{s_m}}}} \right|}^2}\!\!\!<\!\Theta_ 0  \!\!\right) \nonumber \\
&&\!\!\!\!\!\!\!\!\!\!\!\!\!\!\!= \bar P_{\text{o\_{km}}}\left(\!\!\int_0^\infty\!\!\!\!\! \int_{\frac{x_2}{\Delta _1}}^\infty \!\!\!\!\!\! { {{F_U}\!\!\left(\!\! {\Delta _0}\!\!+\!\!2^{\frac{2 R_s}{B}}\!{x_1} \!\!\right)} } \!{f_{X_1}}\!\left(\! x_1 \!\right)\!{f_{X_2}}\!\left(\! x_2 \!\right)d{x_1}d{x_2}\right.\nonumber \\
&&\!\!\!\!\!\!\!\!\!\!\!\!\!\left.+\!\!\! \int_0^\infty\!\!\!\!\! \int_{{\Delta _1}{x_1}}^\infty \!\!\!\!\!\!\!\!\!{ {{F_U}\!\!\left(\!\! {\Delta _0}\!\!+\!\!\frac{2^{\frac{2 R_s}{B}}}{\Delta _1}\!{x_2} \!\!\right)} }\! {f_{X_2}}\!\left(\! x_2 \!\right)\!{f_{X_1}}\!\left(\! x_1 \!\right)d{x_2}d{x_1}\!\!\right)\!\!+\!\!P_{\text{so\_{km}}},
\end{eqnarray}
respectively, where ${F_U(u)}$ is the cumulative distribution function (CDF) of RV $U$, ${f_{X_1}(x_1)}$ and ${f_{X_2}(x_2)}$ are respective the probability density functions (PDFs) of the RVs $X_1$ and $X_2$, $\bar P_{\text{o\_{km}}} = \Pr({{| {{h_{{s_{k}}{s_m}}}} |}^2}>\Theta_ 0)$, and $P_{\text{{so}\_{km}}} = \Pr({{| {{h_{{s_{k}}{s_m}}}} |}^2}<{{2^{\frac{2\cdot R_s}{B}}}} \sum\limits_{l = 1}^{{N_E}}{{{{{| {{h_{{s_{k}}{e_l}}}} |}^2}}}}+\Theta_ 1, {{| {{h_{{s_{k}}{s_m}}}} |}^2}<\Theta_ 0)$. For simplicity, we assume that for different $m, i, j, l, k$, $\sigma _{{s_{{m_i}}}{d_{{m_j}}}}^2 = \sigma _{md}^2$, $\sigma _{{s_{m_i}}{e_l}}^2 = \sigma _{me}^2$, and $\sigma _{{s_k}{e_l}}^2 = \sigma _{me}^2$. Based on [9], they can be expressed as:
\begin{equation}
{F_U}\!\left(\!\! {\Delta _0}\!\!+\!\!2^{\frac{2 R_s}{B}}{x_1} \!\!\right)\!\!=\!\! 1 \!-\! \exp\!\! \left(\!\! { -\! \frac{{\Delta _0}\!\!+\!\!2^{\frac{2 R_s}{B}}{x_1}}{{\sigma _{md}^2}}} \!\!\right)\!\!\!\!\sum\limits_{l = 0}^{{N_T}\!{N_R} \!-\! 1}\!\!\!\!\! {\frac{1}{{l!}}\!{{\left(\!\! {\frac{{\Delta _0}\!\!+\!\!2^{\frac{2 R_s}{B}}{x_1}}{{\sigma _{md}^2}}} \!\!\right)}^l}}
\end{equation}
and
\begin{equation}
{f_{X_1}}\left( {x_1} \right) \!=\! \frac{{{{x_1}^{{N_T}{N_E} - 1}}}}{{\left( {{N_T}{N_E} \!-\! 1} \right)!}}{\left( {\frac{1}{{\sigma _{me}^2}}} \right)^{{N_T}{N_E}}}\!\!\exp \left( { - \frac{{x_1}}{{\sigma _{me}^2}}} \right)
\end{equation}
and
\begin{equation}
{f_{X_2}}\left( {x_2} \right) = \frac{{{{x_2}^{{N_E} - 1}}}}{{\left( {{N_E} - 1} \right)!}}{\left( {\frac{1}{{\sigma _{ke}^2}}} \right)^{{N_E}}}\exp \left( { - \frac{{x_2}}{{\sigma _{ke}^2}}} \right),
\end{equation}
respectively. Substituting (A.3) and (A.4) into (A.1) yields
\begin{eqnarray}
&&\!\!\!\!\!\!\!\!\!\!\!\!\!P_{\text{so\_m\_m}}^{\text{RSDPS}} = \int_0^\infty  { {{F_U}\left( {\Delta _0}+2^{\frac{2 R_s}{B}}{x_1} \right)} } {f_{X_1}}\left( x_1 \right)d{x_1} \nonumber \\
&&\!\!\!\!\!\!\!\!\!\!\!\!\!=\! 1 \!-\!\!\!\!\!\! \sum\limits_{l = 0}^{{N_T}{N_R} - 1} \!\!\!{\sum\limits_{p = 0}^l {\frac{{\left( {p + {N_T}{N_E} - 1} \right)!}}{{p!\left(\! {l - p} \!\right)!\left(\! {{N_T}{N_E} \!-\! 1} \!\right)!}}{{\left(\!\! {\frac{{{2^{{\frac{2 R_s}{B}}}}}}{{\sigma _{md}^2}}} \!\!\right)}^l}} }\!\!{\left(\!\! {\frac{1}{{\sigma _{me}^2}}} \!\!\right)^{{N_T}{N_E}}}\nonumber \\
&&{\left(\!\! {\frac{{{\Delta _0}}}{{{2^{{\frac{2 R_s}{B}}}}}}} \!\!\right)^{l - p}}{\left(\!\! {\frac{1}{{\sigma _{me}^2}} \!\!+\!\! \frac{{{2^{{\frac{2 R_s}{B}}}}}}{{\sigma _{md}^2}}} \!\!\right)^{ - p - {N_T}{N_E}}}\!\!\!\!\!{e^{ - \frac{{{\Delta _0}}}{{\sigma _{md}^2}}}}.
\end{eqnarray}

Relying on [29], the PDF of RV $X_3$ can be approximated as
\begin{equation}
{f_{{X_3}}}\left( x_3 \right) = {{\left(\frac{{m_{{s_k}{s_m}}}}{{\sigma _{{s_k}{s_m}}^2}}\right)}^{m_{{s_k}{s_m}}}}{\frac{{x_3}^{m_B-1}}{\Gamma \left({m_{{s_k}{s_m}}}\right)}}\exp \left( -\frac{ {m _{{s_k}{s_m}}}x_3}{{{\sigma _{{s_k}{s_m}}^2}}} \right),
\end{equation}
where ${m_{{s_k}{s_m}}}={\textstyle{({1 + {K_{{s_k}{s_m}}}})^2 \over {2{K_{{s_k}{s_m}}}+1}}}$, and ${{\sigma _{{s_k}{s_m}}^2}}$ denotes the average power of $|{{h_{{s_k}{s_m}}}}|^2$. Hence, $\bar P_{\text{o\_{km}}}$ and $P_{\text{{so}\_{km}}}$ can be further formulated as
\begin{eqnarray}
&&\!\!\!\!\!\!\!\!\!\!\!\!\!\!\!\!\!\bar P_{\text{o\_{km}}}\!=\!\!\!\int_{\Theta_ 0}^\infty\!\!{{{\left(\!\!\frac{{m_{{s_k}{s_m}}}}{{\sigma _{{s_k}{s_m}}^2}}\!\!\right)}^{m_{{s_k}{s_m}}}}\!\!{\frac{{x_3}^{m_{{s_k}{s_m}}-1}}{\Gamma \left(\!{m_{{s_k}{s_m}}}\!\right)}}\exp\! \left(\!\! -\frac{ {m _{{s_k}{s_m}}}x_3}{{{\sigma _{{s_k}{s_m}}^2}}} \!\!\right)}d{x_3}\nonumber \\
&&\!\!\!=\!\!\!\!\!\!\!\sum\limits_{g = 0}^{m_{{s_k}{s_m}}-1} \!\!\! {\frac{{{{\left( \Theta_ 0\right)}^g}}}{{{{ {g!} }}}}} \exp\! \left( \!\!{ - \frac{\Theta _0{m_{{s_k}{s_m}}}}{{\sigma _{{s_k}{s_m}}^2}}} \!\!\right)\left(\!\frac{{m_{{s_k}{s_m}}}}{{\sigma _{{s_k}{s_m}}^2}}\!\right)^g
\end{eqnarray}
and
\begin{eqnarray}
&&\!\!\!\!\!\!\!\!\!\!\!\!\!P_{\text{{so}\_{km}}}= \left\{ {\begin{array}{*{20}{c}}
{\!\!\!\!\!\!\!\!\!\!\!\!\!\!\!\!\!\!\!\!\!\!\!\!\!\!\!\!\!\!\!\!\!\!\!\!\!\!\!\!\!\!\!\!\!\!\!\!\!\!\!\int_0^{{\Theta _0}} {{f_{{X_3}}}\left( {{x_3}} \right)} d{x_3},\;\; \text{if}\;\;{R_s} \ge {R_o}}\\
{\!\!\!\!\!\!\int_0^{{\Theta _0}} \!\!{{f_{{X_3}}}\left(\! {{x_3}} \!\right)} d{x_3}\int_0^{{\Theta _{{2}}}} \!{{f_{{X_{{2}}}}}\left(\! {{x_{{2}}}} \!\right)} d{x_{{2}}}}\!+\!\! \int_{{0}}^{{\Theta _{{2}}}} \!{\int_0^{{{{2}}^{{\textstyle{{2{R_s}} \over B}}}}{x_2} \!+\! {\Theta _1}}}\\
{\!\!\!\!\!\!\!\!\!\!\!\!\!\!\!\!\!\!\!\!\!\!\!\!\!\!\!\! {{{f_{{X_2}}}\left( {{x_2}} \right){f_{{X_3}}}\left( {{x_3}} \right)} d{x_3}d{x_2}} ,\;\; \text{otherwise}}
\end{array}} \right.\nonumber \\
&&\!\!\!\!\!\!\!\!\!\!\!\!\!= \left\{ {\begin{array}{*{20}{c}}
\!\!\!\!\!\!{1 -\!\!\!\!\!\!\!\! \sum\limits_{g = 0}^{{m_{{s_k}{s_m}}} - 1} \!\!\!\!\!{\frac{{{{\left( {{\Theta _0}} \right)}^g}}}{{g!}}} \exp\! \left(\! { - \frac{{{\Theta _0}{m_{{s_k}{s_m}}}}}{{\sigma _{{s_k}{s_m}}^2}}} \!\right){{\left(\! {\frac{{{m_{{s_k}{s_m}}}}}{{\sigma _{{s_k}{s_m}}^2}}} \!\right)}^g},\;\;{\rm{if}}\;\;{R_s} \ge {R_o}}\\
\!\!\!\!\!\!\!\!\begin{array}{l}
\left(\!\! {1 -\!\!\!\!\!\!\!\!\! \sum\limits_{g = 0}^{{m_{{s_k}{s_m}}} \!-\! 1}\!\!\!\!\!\!\!\! {\frac{{{{\left( {{\Theta _0}} \right)}^g}}}{{g!}}} \!\exp\! \left(\!\! { - \frac{{{\Theta _0}{m_{{s_k}{s_m}}}}}{{\sigma _{{s_k}{s_m}}^2}}} \!\!\right)\!{{\left(\!\! {\frac{{{m_{{s_k}{s_m}}}}}{{\sigma _{{s_k}{s_m}}^2}}} \!\!\right)}^g}} \!\!\right)\left(\!\! {\sum\limits_{g = 0}^{{N_E} - 1} \!\!\!\!{\frac{1}{{g!}}} {{\left( {\frac{{{\Theta _2}}}{{\sigma _{me}^2}}}\right.}}}\right. \\
\left.{{{\left.\exp\!\! \left(\!\! { - \frac{{{\Theta _2}}}{{\sigma _{me}^2}}} \!\!\right)\!\!\right)}^g}} \!\!\right)\!\!+\!\! {\Theta _3}{\Theta _4}\!\left(\!\! {\left(\! {{N_E} \!-\! 1} \!\right)!}{{\left(\!\! {\frac{1}{{\sigma _{me}^2}}} \!\!\right)}^{ - {N_E}}} \!\!\!\!\!-\! \exp\! \left(\!\! { - \frac{{{\Theta _2}}}{{\sigma _{me}^2}}} \!\!\right)\right.\\
\left.{\sum\limits_{g = 0}^{{N_E} - 1} \!\!\!{\frac{{\left(\! {{N_E} - 1}\! \right)!}}{{g!}}} {{\left( {{\Theta _2}} \right)}^g}{{\left(\!\! {\frac{1}{{\sigma _{me}^2}}} \!\!\right)}^{{N_E} - g}}} \!\!\right) \!-\!\!\!\!\! \sum\limits_{g = 0}^{{m_{{s_k}{s_m}}} \!-\! 1}\!\!\! {\sum\limits_{p = 0}^g \!\!\!{\frac{{\left( {{N_E} - 1} \right)!{\Theta _4}}}{{p!\left( {g - p} \right)!}}}}\\
{{{\left(\!\! {\frac{{{m_{{s_k}{s_m}}}}}{{\sigma _{{s_k}{s_m}}^2}}} \!\!\right)}^{ - {m_{{s_k}{s_m}}} \!+\! g}}\!\!{{{\left(\!\! {\frac{{{2^{\frac{{2{R_s}}}{B}}} \!-\! 1}}{{{\gamma _s}{2^{\frac{{2{R_s}}}{B}}}}}} \!\!\right)}^{g \!-\! p}}} }\!\!\!\!\!\!\!\!\exp\!\! \left(\!\! { - \frac{{\left(\!\! {{2^{\frac{{2{R_s}}}{B}}} \!-\! 1} \!\!\right){m_{{s_k}{s_m}}}}}{{{\gamma _s}\sigma _{{s_k}{s_m}}^2}}} \!\!\right) \\
\left(\!\! {\frac{{{\left(\!\! {{2^{\frac{{2{R_s}}}{B}}}} \!\!\right)}^g}{\left( {p + {N_E} - 1} \right)!}}{{{{\left( {\frac{1}{{\sigma _{me}^2}} + {2^{\frac{{2{R_s}}}{B}}}\frac{{{m_{{s_k}{s_m}}}}}{{\sigma _{{s_k}{s_m}}^2}}} \!\!\right)}^{p \!+\! {N_E}}}}} \!-\!\!\!\!\!\!\! \sum\limits_{l = 0}^{p \!+\! {N_E} \!-\! 1}\!\!\!\!\! {\frac{{\left( {p + {N_E} - 1} \right)!{{\left( {{\Theta _2}} \right)}^l}}}{{l!{{\left(\!\! {\frac{1}{{\sigma _{me}^2}} \!+ \! {2^{\frac{{2{R_s}}}{B}}}\frac{{{m_{{s_k}{s_m}}}}}{{\sigma _{{s_k}{s_m}}^2}}} \!\!\right)}^{p \!+\! {N_E} \!-\! l}}}}}}\right.\\
\left.{{\exp\!\! \left(\!\! { - {\Theta _2}\!\left(\! {\frac{1}{{\sigma _{me}^2}} \!+\! {2^{\frac{{2{R_s}}}{B}}}\frac{{{m_{{s_k}{s_m}}}}}{{\sigma _{{s_k}{s_m}}^2}}} \right)} \!\!\right)} } \!\!\right),{\rm{otherwise}}
\end{array}
\end{array}} \right.,
\end{eqnarray}
respectively, where $\Theta _2 \!\!=\!\! \frac{1}{{{2^{\frac{2 {R_s}}{B}}}}}(\Theta _0\!-\!\Theta _1)$, $\Theta _3\!\!=\!\!\frac{(m_{{s_k}{s_m}}-1)!{\sigma _{{s_k}{s_m}}^{2{m_{{s_k}{s_m}}}}}}{(m_{{s_k}{s_m}})^{m_{{s_k}{s_m}}}}$, and $\Theta _4\!\!=\!\!\frac{1}{(N_E-1)!\Gamma({m_{{s_k}{s_m}}})}{(\frac{1}{\sigma _{me}^2})}^{N_E}(\frac{{m_{{s_k}{s_m}}}}{{\sigma _{{s_k}{s_m}}^2}})^{{m_{{s_k}{s_m}}}}$. Furthermore, substituting (A.3)-(A.5), and (A.8)-(A.9) into (A.2) yields
\begin{eqnarray}
&&\!\!\!\!\!\!\!\!\!\!\!\!\!\!\!\!P_{\text{so\_k\_m}}^{\text{RSDPS}} = \!\!\bar P_{\text{o\_{km}}}\left(\!\!\sum\limits_{t = 0}^{{N_T}{N_E} \!-\! 1}\!\!\!\! {{\left(\!\!\frac{1}{\sigma _{ke}^2}\!\!\right)^{N_E}\!\!\!\!\left(\!\!\frac{1}{{\sigma _{me}^2}{\Delta _1}}\!\!\right)^{t} \!\!\frac{\left(\!t\!+\!N_E\!-\!1\!\right)!}{t!\left(\!N_E\!-\!1\!\right)!} } {{ {{c_{km}^{ - t\!-\! {N_E}}}} }}}\right.\nonumber \\
&&\!\!\!\!\!\!\left.-\!\!\!\!\!\! \sum\limits_{l = 0}^{{N_T}{N_R} - 1}\!\!\sum\limits_{p = 0}^{l}\!\!\sum\limits_{t = 0}^{p\!+\!{N_T}{N_E} \!-\! 1}\!\!\!\!\!\! {{a_{lp}{c_{md}}}{{ {{{\left(\!c_{km}\!+\!\frac{2^{\frac{2 R_s}{B}}}{{\Delta _1}{\sigma _{md}^2}}\!\right)}^{ - t- {N_E}}}} }}} \right. \nonumber \\
&&\!\!\!\!\!\! +\left.\!\!\!\sum\limits_{t = 0}^{{N_E} - 1}\!\!\! {{\left(\!\frac{1}{\sigma _{me}^2}\!\right)^{{N_T}{N_E}}\!\!\!\left(\!\frac{{\Delta _1}}{{\sigma _{ke}^2}}\!\right)^{t} \frac{\left(\!t\!+{\!N_T}{\!N_E}\!-\!1\!\right)!}{t!\left({\!N_T}{\!N_E\!}-\!1\!\right)!} } {{ {{d_{km}^{ - t\!-\! {N_T}{N_E}}}} }}}\right.\nonumber \\
&&\!\!\!\!\!\!\left. -\!\!\!\!\!\!\!\! \sum\limits_{l = 0}^{{N_T}{N_R} \!-\! 1}\!\!\!\!\sum\limits_{p = 0}^{l}\!\!\!\!\sum\limits_{t = 0}^{p\!+\!{N_E} \!-\! 1}\!\!\!\!\!\!\! {{a_{lp}{d_{kd}}}{{ {{{\left(\!\!d_{km}\!\!+\!\!\frac{2^{\frac{2 R_s}{B}}}{{\sigma _{md}^2}}\!\!\right)}^{ - t- {N_T}{N_E}}}} }}}\!\!\right)\!\!+\!\!P_{\text{{so}\_{km}}},
\end{eqnarray}
where $a_{lp}\!=\!\frac{{(\! {\frac{1}{{\sigma _{ke}^2}}} \!)^{N_E}}{(\! {\frac{1}{{\sigma _{me}^2}}} \!)^{{N_T}{N_E}}}{(\! {\frac{2^{\frac{2 R_s}{B}}}{{\sigma _{md}^2}}} \!)^l}{(\! {\frac{{{\Delta _0}}}{{{2^{{\frac{2 R_s}{B}}}}}}} \!)^{l-p}} e^{-\frac{\Delta _0}{\sigma _{md}^2}}}{p!(l-p)!t!(N_E-1)!({N_T}{N_E}-1)!}$, $c_{md}\!\! =\!\! ({ \frac{1}{{\sigma _{me}^2}} \!\!+\!\! \frac{2^{\frac{2 R_s}{B}}}{{\sigma _{md}^2}}})^{-p-{N_T}\!{N_E}+t}{\Delta _1}^{-t}(p\!+\!{N_T}{N_E}\!-\!1)!(t\!+\!N_E\!-\!1)!$, $d_{kd} = ({ \frac{1}{{\sigma _{ke}^2}} + \frac{2^{\frac{2 R_s}{B}}}{{\Delta _1}{\sigma _{md}^2}}})^{-p-{N_E}+t}{\Delta _1}^{t-p}(t+{N_T}{N_E}-1)!(p+N_E-1)!$, $c_{km} = { \frac{1}{{\sigma _{ke}^2}} + \frac{1}{{\Delta _1}{\sigma _{me}^2}}}$, and $d_{km} = { \frac{{\Delta _1}}{{\sigma _{ke}^2}} + \frac{1}{{\sigma _{me}^2}}}$.

Moreover, defining $Q={{\sum\limits_{j = 1}^{{N_R}} {{{{{| {{h_{{s_{{m_i}}}{d_{{m_j}}}}}} |}^2}}}} }}$, ${W_1}={\sum\limits_{l = 1}^{{N_E}} {{{{{| {{h_{{s_{{m_i}}}{e_l}}}} |}^2}}}} }$, and ${W_2}={\sum\limits_{l = 1}^{{N_E}} {{{{{| {{h_{{s_{k}}{e_l}}}} |}^2}}}} }$, and exploiting that the RVs $Q$, $W_1$ and $W_2$ are independent of each other, $P_{\text{so\_s}}^{\text{TAS}}$ and $P_{\text{so\_k}}^{\text{TAS}}$ can be formulated as
\begin{eqnarray}
&&\!\!\!\!\!\!\!\!\!\!\!\!\!\!P_{\text{so\_s}}^{\text{TAS}} \!=\! {\Pr\! \left(\!\! {\mathop {\max }\limits_{m \in \mathbb{D},1 \le i \le {N_T}}\!\!{{\sum\limits_{j = 1}^{{N_R}} \!{{{{{\left| {{h_{{s_{{m_i}}}{d_{{m_j}}}}}} \right|}^2}}}} }}  \!\!<\!\!2^{\frac{2 R_s}{B}}{\sum\limits_{l = 1}^{{N_E}}\! {{{{{\left|\! {{h_{{s_{{m_i}}}{e_l}}}} \!\right|}^2}}}} }\!+\!\Lambda _0  }\!\!\right)}\nonumber \\
&&\!= \int_0^\infty \!\!\!\!\! {\prod\limits_{m \in \mathbb{D},1 \le i \le {N_T}} \!\!\!\!{{F_Q}\left( {\Lambda _0}+2^{\frac{2 R_s}{B}}{w_1} \right)} } {f_{W_1}}\left( w_1 \right)d{w_1}
\end{eqnarray}
and
\begin{eqnarray}
&&\!\!\!\!\!\!\!\!\!\!\!\!\!\!\!\!\!\!P_{\text{so\_k}}^{\text{TAS}} \!\!=\!\! \Pr\!\! \left(\!\! \mathop {\max }\limits_{m \in \mathbb{D},1 \le i \le {N_T}}{ {\sum\limits_{j = 1}^{{N_R}} {{{\left| {{h_{{s_{{m_i}}}{d_{{m_j}}}}}} \right|}^2}} }  \!\!\!\!<}{{{\max}\left(\!\!{2^{{\frac{2 R_s}{B}}}}\!{\sum\limits_{l = 1}^{{N_E}}\! {{{{{\left|\! {{h_{{s_{{m_i}}}{e_l}}}} \!\right|}^2}}}} }\!,\right.}}\right.\nonumber \\
&&\;\;\;\;\;\left.{{\left.{\frac{{2^{{\frac{2 R_s}{B}}}}}{{{\Lambda _1}}}}\!{\sum\limits_{l = 1}^{{N_E}} \!{{{{{\left|\! {{h_{{s_{k}}{e_l}}}} \!\right|}^2}}}} }\!\!\right) \!\!+\!\! {\Lambda _0}}} \!\!\right)\bar P_{\text{{o}\_{km}}}\!+\!P_{\text{{so}\_{km}}}\nonumber \\
&&\!\!\!\!\!\!\!\!\!\!\!\!\!\!\!\!\!\!=\!\! \left(\!\!\int_0^\infty \!\!\!\int_{\frac{w_2}{\Lambda _1}}^\infty \!\!\!{\prod\limits_{m \in \mathbb{D},1 \le i \le {N_T}}\!\!\!\!\!\!\!\!\!\!\! { {{F_Q}\!\!\left(\!\! {\Lambda _0}\!\!+\!\!2^{\frac{2 R_s}{B}}{w_1} \!\!\right)} }} {f_{W_1}}\!\left(\! w_1 \!\right){f_{W_2}}\!\left(\! w_2 \!\right)d{w_1}d{w_2}\right.\nonumber \\
&&\!\!\!\!\!\!\!\!\!\!\!\!\!\!\!\!\!+\!\!\!\int_0^\infty \!\!\! \int_{{\Lambda _1}{w_1}}^\infty\!\!\! {\prod\limits_{m \in \mathbb{D},1 \le i \le {N_T}} \!\!\!\!\!\!\!{ {{F_Q}\!\left(\!\! {\Lambda _0}\!\!+\!\!\frac{{2^{\frac{2 R_s}{B}}}{w_2}}{\Lambda _1} \!\right)} }} {f_{W_2}}\left( w_2 \right){f_{W_1}}\left( w_1 \right)\nonumber \\
&&\;\;\;\;d{w_2}d{w_1}\bigg)\bar P_{\text{{o}\_{km}}}\!+\!P_{\text{{so}\_{km}}},
\end{eqnarray}
respectively.

Based on [9], ${F_Q}(w)$, ${f_{W_1}}({w_1})$ and ${f_{W_2}}({w_2})$ can be formulated as:
\begin{equation}
{F_Q}\left( w \right) = 1 - \exp \left( { - \frac{w}{{\sigma _{md}^2}}} \right)\sum\limits_{l = 0}^{{N_R} - 1} {\frac{1}{{l!}}{{\left( {\frac{w}{{\sigma _{md}^2}}} \right)}^l}}
\end{equation}
and
\begin{equation}
{f_{W_1}}\left( {w_1} \right) = \frac{{{{w_1}^{{N_E} - 1}}}}{{\left( {{N_E} - 1} \right)!}}{\left( {\frac{1}{{\sigma _{me}^2}}} \right)^{{N_E}}}\exp \left( { - \frac{{w_1}}{{\sigma _{me}^2}}} \right)
\end{equation}
and
\begin{equation}
{f_{W_2}}\left( {w_2} \right) = \frac{{{{w_2}^{{N_E} - 1}}}}{{\left( {{N_E} - 1} \right)!}}{\left( {\frac{1}{{\sigma _{ke}^2}}} \right)^{{N_E}}}\exp \left( { - \frac{{w_2}}{{\sigma _{ke}^2}}} \right),
\end{equation}
respectively. Substituting (A.17) and (A.18) into (A.15) yields
\begin{eqnarray}
&&\!\!\!\!\!\!\!\!\!\!\!\!\!\!\!\!P_{\text{so\_s}}^{\text{TAS}}\!=\!\!\!\! \int_0^\infty  \!\!\!{\prod\limits_{m \in \mathbb{D},1 \le i \le {N_T}} \!\!\!{{F_Q}\left( {\Lambda _0}+2^{\frac{2 R_s}{B}}w \right)} } {f_W}\left( w \right)dw \nonumber \\
&&\!\!\!\!\!\!\!\!\!\!\!\!\!\!\!\!= \sum\limits_{S'} {\sum\limits_{p = 0}^{{\beta _2}} {\int_0^\infty  {{\Psi _0}{w^{p + {N_E} - 1}}\exp \left( { - \frac{w}{{\sigma _{me}^2}} - {\beta _3}{2^{{\frac{2 R_s}{B}}}}w} \right)} dw} }  \nonumber \\
&&\!\!\!\!\!\!\!\!\!\!\!\!\!\!\!\! =\!\! \sum\limits_{S'} \!\!{\sum\limits_{p = 0}^{{\beta _2}} {{\Psi _0}\left(\! {p + {N_E} - 1} \!\right)!{{\left(\! {\frac{1}{{\sigma _{me}^2}} + {\beta _3}{2^{{\frac{2 R_s}{B}}}}} \!\right)}^{ - p - {N_E}}}} } ,
\end{eqnarray}
where ${\beta _1} \!=\! \frac{{( {| \mathbb{D} |}{\cdot}{{N_T} } )!}}{{\prod\limits_{i = 1}^{{{N_R}} + 1} \!\!{{n_i}!} }}\!\!\prod\limits_{j = 1}^{{{N_R}}} {{{( { - \frac{1}{{{\sigma _{md}^{2( {j - 1} )}}( {j - 1} )!}}} )}^{{n_j}}}}$, ${\beta _2} \!=\!\!\! \sum\limits_{j = 1}^{{{N_R}}}\!\! {{n_j}( {j - 1} )} $, ${S'} = \{ { {( {{n_1},{n_2}, \cdots \!,\!{n_{{{N_R}} + 1}}} )} |\!\!\sum\limits_{i = 1}^{{{N_R}} + 1} \!\!{{n_i}}  \!=\! | \mathbb{D}| \!\cdot\!{N_T}} \}$, ${\beta _3} = \frac{1}{{\sigma _{md}^2}}( {| \mathbb{D} | \cdot {N_T}- {n_{{{N_R}} + 1}}} )$, and ${\Psi _0} \!=\! \frac{{{\beta _1}}}{{( {{N_E} - 1} )!}} ( \begin{array}{l}{\beta _{\rm{2}}}\\p\end{array}){( {\frac{1}{{\sigma _{me}^2}}} )^{{N_E}}}{( {{2^{{\frac{2 R_s}{B}}}}} )^{{\beta _2}}}{( {\frac{{{\Lambda _0}}}{{{2^{{\frac{2 R_s}{B}}}}}}} )^{{\beta _2} - p}}{e^{ - {\beta _3}{\Lambda _0}}}$.

Using (A.17)-(A.19), and (A.8)-(A.9), we arrive at
\begin{eqnarray}
&&\!\!\!\!\!\!\!\!\!\!\!\!P_{\text{so\_k}}^{\text{TAS}} \!\!=\!\!\bar P_{\text{o\_{km}}}\left(\!\! {\sum\limits_S {\sum\limits_{p = 0}^{\beta _2}\!\! \sum\limits_{t = 0}^{p \!+\! {N_E} \!-\! 1}\!\!\!\!\!{a_{{\beta}{p}}{c_{{\beta}d}}{{\left(\!\! \frac{d_{km}^{'}}{\Lambda _1} \!+\! \frac{{\beta _3}{2^{\frac{2 R_s}{B}}}}{\Lambda _1} \!\!\right)}^{ - t \!-\! {N_E}}}} } }   \right.\nonumber \\
&&\!\!\!\!\!\!\!\!\!\!\!\!+\!\!\!\left.{\sum\limits_S\! {\sum\limits_{p = 0}^{\beta _2}\!\!\sum\limits_{t = 0}^{p\!+\! {N_E} \!-\! 1}\!\!\!\!\! {a_{{\beta}{p}}{d_{{\beta}d}}}}}{{{{{\left(\!\! {c_{km}^{'}}{\Lambda _1} \!\!+\!\! {{\beta _3}{2^{\frac{2 R_s}{B}}}} \!\!\right)}^{ - t \!-\! {N_E}}}} } }\!\!\right) \!\!+\!\!P_{\text{{so}\_{km}}},
\end{eqnarray}
where $a_{{\beta}p}=\frac{{( {\frac{1}{{\sigma _{ke}^2}}} )^{N_E}}{( {\frac{1}{{\sigma _{me}^2}}} )^{{N_E}}}{( {{2^{\frac{2 R_s}{B}}}} )^p}{( {{{{\Lambda _0}}}} )^{{\beta _2}-p}}{\beta _1}(\beta _2)! e^{-{\Lambda _0}{\beta _3}}}{p!({\beta _2}-p)!t!(N_E-1)!({N_E}-1)!}$, $c_{{\beta}d} = ({ \frac{1}{{\sigma _{me}^2}} + {2^{\frac{2 R_s}{B}}}{{\beta _3}}})^{-p-{N_E}+t}{\Lambda _1}^{-t}(p+{N_E}-1)!(t+N_E-1)!$, $c_{km}^{'} \!=\! { \frac{1}{{\sigma _{ke}^2}} \!+\! \frac{1}{{\Lambda _1}{\sigma _{me}^2}}}$, $d_{km}^{'} \!=\! { \frac{{\Lambda _1}}{{\sigma _{ke}^2}} \!+\! \frac{1}{{\sigma _{me}^2}}}$, and $d_{{\beta}d} \!=\! ({ \frac{1}{{\sigma _{ke}^2}} \!+\! \frac{{2^{\frac{2 R_s}{B}}}{\beta _3}}{{\Lambda _1}}})^{-p-{N_E}+t}{\Lambda _1}^{t-p}(t+{N_E}-1)!(p+N_E-1)!$.

\section{}
\emph{A, Proof of Theorem 1:}

Upon utilizing (18), (19), and the inequality $\sum\limits_{i = 1}^{{N_T}} {\sum\limits_{j = 1}^{{N_R}} {{{{{| {{h_{{s_{{m_i}}}{d_{{m_j}}}}}} |}^2}}} }  \le {{{N_T}{N_R}}}\mathop {\max }\limits_{i,j} {| {{h_{{s_{{m_i}}}{d_{{m_j}}}}}} |^2}}$, ${2^{\frac{2 R_s}{B}}}\!\!\sum\limits_{i = 1}^{{N_T}} {\sum\limits_{l = 1}^{{N_E}} {{{{{| {{h_{{s_{{m_i}}}{e_l}}}} |}^2}}}} }\!\!+\!\!\Delta _0  \!\!\ge\!\! {2^{\frac{2 R_s}{B}}}\!\!\mathop {\max }\limits_{i,l}\! {| {{h_{{s_{{m_i}}}{e_l}}}} |^2}\!\!$, and ${2^{\frac{2 R_s}{B}}}{\max}{(\!\sum\limits_{i = 1}^{{N_T}} \!{\sum\limits_{l = 1}^{{N_E}}\!\! {{{{{| {{h_{{s_{{m_i}}}{e_l}}}} |}^2}}}} }\!,\!\frac{1}{\Delta _1} \!\!{\sum\limits_{l = 1}^{{N_E}}\!\! {{{{{| {{h_{{s_{{m_i}}}{e_l}}}} |}^2}}}} })}\!+\!\Delta _0  \!\!\ge\!\! {2^{\frac{2 R_s}{B}}}{\max}{(\mathop {\max }\limits_{i,l} {| {{h_{{s_{{m_i}}}{e_l}}}} |^2}\!,\!\frac{1}{\Delta _1}\mathop {\max }\limits_{l} {| {{h_{{s_{k}}{e_l}}}} |^2})}$, we have
\begin{eqnarray}
&&\!\!\!\!\!\!\!\!\!\!\!\!\!\!\!\! P_{\text{so}}^{\text{RSDPS}} \!\!\ge\!\! \frac{1}{M}\!\!\!\!\sum\limits_{m = 1}^M \!\!\!\!{\frac{1}{M}\!\!\left(\!\!\Pr\! {\left(\!\! {{{{N_T}{N_R}}}\mathop {\max }\limits_{i,j} {{\left|\! {{h_{{s_{{m_i}}}{d_{{m_j}}}}}} \!\right|}^2} \!\!\!\!< \!}\right.}{{2^{\frac{2 R_s}{B}}}\mathop {\max }\limits_{i,l} {{\left|\! {{h_{{s_{{m_i}}}{e_l}}}} \!\right|}^2}} \!\!\right)} \nonumber \\
&&+\!\!\!\!\!\!\!\!\sum\limits_{k \in \mathbb{D} - \left\{ m \right\}}^{}\!\!\!\!\!\Pr\!\! \left({\!\! {{{{N_T}{N_R}}}\mathop {\max }\limits_{i,j} {{\left|\! {{h_{{s_{{m_i}}}{d_{{m_j}}}}}} \!\right|}^2}\!\!\!\! <\! }{{2^{\frac{2 R_s}{B}}}{\max}\!\!\left(\!{\mathop {\max }\limits_{i,l} {{\left| {{h_{{s_{{m_i}}}{e_l}}}} \right|}^2}},\right.}}\right.\nonumber \\
&&\;\;\;\;\;\;\;\;\;\;\left.{\left.{\left.\frac{1}{\Delta _1}{\mathop {\max }\limits_{l} {{\left| {{h_{{s_{k}}{e_l}}}} \right|}^2}}\!\right)} \!\!\right)}\!\!\right){\bar P_{\text{o\_{km}}}}.
\end{eqnarray}

Defining ${X_1} \!=\! \mathop {\max }\limits_{i,l} {| {{h_{{s_{{m_i}}}{e_l}}}} |^2}$, ${X_2} \!=\! \mathop {\max }\limits_{i,l} {| {{h_{{s_{k}}{e_l}}}} |^2}$, and $Y\! =\! \mathop {\max }\limits_{i,l} {| {{h_{{s_{{m_i}}}{d_{{m_j}}}}}} |^2}$, the expressions $\Pr ( {\mathop {\max }\limits_{i,j} {{| {{h_{{s_{{m_i}}}{d_{{m_j}}}}}} |}^2} <{2^{\frac{2 R_s}{B}}}}$\\${ \frac{1}{{{{N_T}{N_R}}}}\mathop {\max }\limits_{i,l} {{| {{h_{{s_{{m_i}}}{e_l}}}} |}^2}} )$ and $\Pr ( {\mathop {\max }\limits_{i,j} {{| {{h_{{s_{{m_i}}}{d_{{m_j}}}}}} |}^2} <{2^{\frac{2 R_s}{B}}}\frac{1}{{{{N_T}{N_R}}}}}$\\${{{\max}(\mathop {\max }\limits_{i,l} {{| {{h_{{s_{{m_i}}}{e_l}}}} |}^2},\frac{1}{{\Delta _1}}\mathop {\max }\limits_{l} {{| {{h_{{s_{k}}{e_l}}}} |}^2})}} )$ can be rewritten as
\begin{eqnarray}
&&\Pr \left( {\mathop {\max }\limits_{i,j} {{\left| {{h_{{s_{{m_i}}}{d_{{m_j}}}}}} \right|}^2}} < \frac{{2^{\frac{2 R_s}{B}}}}{{{{N_T}{N_R}}}}\mathop {\max }\limits_{i,l} {{\left| {{h_{{s_{{m_i}}}{e_l}}}} \right|}^2} \right)\nonumber \\
&&= \int_0^\infty {{{\prod\limits_{i,j}^{}{F_Y} \left(\frac{ {{2^{\frac{2 R_s}{B}}} {x_1}}}{{{N_T}{N_R}}} \right)}}{f_{X_1}}\left( {x_1} \right)d{x_1}}
\end{eqnarray}
and
\begin{eqnarray}
&&\!\!\!\!\!\!\!\!\!\!\!\!\!\!\!\Pr\! \left(\!\! {\mathop {\max }\limits_{i,j} \!{{\left|\! {{h_{{s_{{m_i}}}{d_{{m_j}}}}}} \!\right|}^2} \!\!\!\!\!<\! }{\frac{2^{\frac{2 R_s}{B}}}{{{N_T}{N_R}}}{\max}\!\!\left(\!\!\mathop {\max }\limits_{i,l}\! {{\left|\! {{h_{{s_{{m_i}}}{e_l}}}} \!\right|}^2}\!,\!\frac{1}{\Delta _1}\mathop {\max }\limits_{l}\! {{\left|\! {{h_{{s_{k}}{e_l}}}} \!\right|}^2}\!\!\right)} \!\!\right) \nonumber \\
&&\!\!\!\!\!\!\!\!\!\!\!\!\!\!\!= \int_0^\infty\!\!\!\! \int_{\frac{x_2}{\Delta _1}}^\infty\!\!  {{\prod\limits_{i,j}^{}{F_Y} \left( \frac{{2^{\frac{2 R_s}{B}}} {x_1}}{{N_T}{N_R}} \right)}{f_{X_1}}\left(\! x_1 \!\right){f_{X_2}}\left(\! x_2 \!\right)d{x_1}d{x_2}} \nonumber \\
&&\!\!\!\!\!\!\!\!\!\!+  \int_0^\infty\!\!\!\! \int_{{\Delta _1}{x_1}}^\infty \!\!\!  {{\prod\limits_{i,j}^{}\!\!{F_Y}\!\! \left(\!\! \frac{{2^{\frac{2 R_s}{B}}} {x_2}}{{N_T}{N_R}{\Delta _1}} \!\!\right)}{f_{X_2}}\left(\! x_2 \!\right){f_{X_1}}\left(\! x_1 \!\right)d{x_2}d{x_1}},
\end{eqnarray}
respectively, where ${{F_Y}{(y)}}$ is the CDF of the RV $Y$, while ${{f_{X_1}}{({x_1})}}$ and ${{f_{X_2}}{({x_2})}}$ are the PDFs of the RVs $X_1$ and $X_2$, respectively.

Noting that the RVs ${| {{h_{{s_{{m_i}}}{e_l}}}} |^2}$ and ${| {{h_{{s_{k}}{e_l}}}} |^2}$ obey the exponential distribution and are independent of each other, $i = 1,2, \cdots ,{N_T}$, $l = 1,2, \cdots ,{N_E}$, the CDF of $X_1$ can be expressed as:
\begin{eqnarray}
&&\!\!\!\!\!\!\!\!\!\!\!\!\Pr\! \left( {X \!<\! x} \right) \!=\! \Pr\! \left(\!\! {\mathop {\max }\limits_{i,l} {{\left| {{h_{{s_{{m_i}}}{e_l}}}} \right|}^2} \!< \!x} \!\!\right) = \prod\limits_{i,l} {\Pr \left( {{{\left| {{h_{{s_{{m_i}}}{e_l}}}} \right|}^2} < x} \right)} \nonumber \\
&&\!\!\!\!\!\!\!\!\!\!\!\!= 1 +\!\!\!\!\!\! \sum\limits_{n = 1}^{{2^{{N_T}{N_E}}} - 1} \!\!\!\!\!\!\!{{{\left( { - 1} \right)}^{\left| {{C_n}} \right|}}\exp \left(\!\! { - \!\!\!\!\sum\limits_{i,l \in {C_n}}^{} {\frac{x}{{\sigma _{{s_{m_i}}{e_l}}^2}}} } \!\!\right)},
\end{eqnarray}
where $| {{C_n}} |$ is the cardinality of the set ${C_n}$, and ${C_n}$ denotes the $n$-th non-empty subset of ${C}$. Moreover, $C$ represents the set of the links spanning from a SN to the eavesdropper E in the second stage.

Hence, the PDF of the RV $X_1$ can be formulated as
\begin{equation}
{f_{X_1}}\left( {x_1} \right) = \sum\limits_{n = 1}^{{2^{{N_T}{N_E}}} - 1} \!{\sum\limits_{i,l \in {C_n}}^{} {\frac{{{{\left(\! { - 1} \!\right)}^{\left| {{C_n}} \right| + 1}}}}{{\sigma _{{s_{m_i}}{e_l}}^2}}} \exp\! \left(\! { -\! \sum\limits_{i,l \in {C_n}}^{} \!{\frac{{x_1}}{{\sigma _{{s_{m_i}}{e_l}}^2}}} } \!\right)}.
\end{equation}

Similarly to (B.5), the PDF of the RV $X_2$ is given by
\begin{equation}
{f_{X_2}}\left( {x_2} \right) = \sum\limits_{n = 1}^{{2^{{N_E}}} - 1} \!{\sum\limits_{l \in {F_g}}^{} \!{\frac{{{{\left(\! { - 1} \!\right)}^{\left| {{F_g}} \right| + 1}}}}{{\sigma _{{s_k}{e_l}}^2}}} \exp\! \left(\! { -\! \sum\limits_{l \in {F_g}}^{}\! {\frac{{x_2}}{{\sigma _{{s_k}{e_l}}^2}}} } \!\right)},
\end{equation}
where $| {{F_g}} |$ represents the cardinality of the set ${F_g}$, and ${F_g}$ is the $g$-th non-empty subset of ${F}$. Moreover, $F$ denotes the set of the links spanning from a SN to the eavesdropper E in the first stage. Furthermore, $ \prod\limits_{i,j}^{}{{F_Y}(\frac{ {{2^{\frac{2 R_s}{B}}} {x_1}}}{{{N_T}{N_R}}})}$ can be expanded as
\begin{equation}
 \prod\limits_{i,j}^{}\!{{F_Y}\!\!\left(\! \frac{{2^{\frac{2 R_s}{B}}{x_1}}}{{N_T}{N_R}} \!\right)} \!\!=\prod\limits_{i,j}^{} {\left(\! {1 \!-\! \exp \!\left(\! { - \frac{2^{\frac{2 R_s}{B}}}{{ {N_T}{N_R}}}\frac{{x_1}}{{\sigma _{{s_{{m_i}}}{d_{{m_j}}}}^2}}} \!\right)} \!\right)}.
\end{equation}

For notational convenience, we introduce $Z_1 = { - \frac{2^{\frac{2 R_s}{B}}}{{ {N_T}{N_R}}}\frac{{x_1}}{{\sigma _{{s_{{m_i}}}{d_{{m_j}}}}^2}}}$, and $Z_2 = { - \frac{2^{\frac{2 R_s}{B}}}{{ {N_T}{N_R}}}\frac{{x_2}}{{\Delta _1}{\sigma _{{s_{{m_i}}}{d_{{m_j}}}}^2}}}$. Then, $E({Z_1})$ is given by
\begin{eqnarray}
&&\!\!\!\!\!\!\!\!\!\!\!\!E \left({Z_1}\right) \!=\!\!\! \int_0^\infty\!\!\!\! \int_{\frac{x_2}{\Delta _1}}^\infty\!\!  {\left(\!\!\frac{2^{\frac{2 R_s}{B}}}{{ {N_T}{N_R}}}\frac{{x_1}}{{\sigma _{{s_{{m_i}}}{d_{{m_j}}}}^2}}\!\!\right){f_{X_1}}\left(\! x_1 \!\right){f_{X_2}}\left(\! x_2 \!\right)d{x_1}d{x_2}} \nonumber \\
&&\!\!\!\!\!\!\!\!\!\!\!\!=\!\!\!\!\!\!\sum\limits_{n = 1}^{{2^{{N_T}{N_E}}} \!-\! 1}\!\sum\limits_{g = 1}^{{2^{{N_E}}} \!-\! 1}\!\sum\limits_{t = 0}^{1}{\left(\frac{1}{\Delta _1}\right)^t} \frac{{2^{\frac{2 R_s}{B}}}{a_{ngt}}{\left(-1\right)^{\left|C_n\right|+\left|F_g\right|}}}{{{N_T}{N_R}}\left(1-t\right)!}\frac{1}{{\lambda _{se}}},
\end{eqnarray}
where $a_{ngt}=\frac{{(\!\!\! {\sum\limits_{i,l \in {C_n}}^{} \!\!\!\!\!{\frac{1}{{\alpha _{{s_{{m_i}}}{e_l}}^{}}}} } \!)^{t \!-\! 1}}(\!\! {\sum\limits_{l \in {F_g}}^{} \!\!\!\!{\frac{1}{{\alpha _{{s_k}{e_l}}^{}}}} } \!){(\! { -\!\!\!\!\!\! \sum\limits_{i,l \in {C_n}}^{}\!\!\!\! {\frac{1}{{\alpha _{{s_{{m_i}}}{e_l}}^{}}}}  \!-\!\!\!\! \sum\limits_{l \in {F_g}}^{} {\frac{1}{{\alpha _{{s_k}{e_l}}^{}}}} } \!)^{ - t \!-\! 1}}}{{\alpha _{{s_{{m_i}}}{d_{{m_j}}}}^{}}}$. Upon considering ${\lambda _{se}} \to \infty $, $E({Z_1})$ tends to zero. Similarly, $E({Z_2})$, $E(({Z_1})^2)$ and $E(({Z_2})^2)$ also tend to zero, when ${\lambda _{se}} \to \infty $.
Thus, based on [25], $1 - \exp ( { - \frac{1}{{{N_T}{N_R}}}\frac{{2^{\frac{2 R_s}{B}}x}}{{\sigma _{{s_{{m_i}}}{d_{{m_j}}}}^2}}} )$ can be simplified to
\begin{equation}
1 - \exp \left( { - \frac{1}{{ {N_T}{N_R}}}\frac{{2^{\frac{2 R_s}{B}}x}}{{\sigma _{{s_{{m_i}}}{d_{{m_j}}}}^2}}} \right)\mathop  = \limits^1 \frac{{2^{\frac{2 R_s}{B}}}}{{ {N_T}{N_R}}}\frac{x}{{\sigma _{{s_{{m_i}}}{d_{{m_j}}}}^2}}.
\end{equation}

Hence, $ \prod\limits_{i,j}^{}{{F_Y}(\frac{ {{2^{\frac{2 R_s}{B}}} {x_1}}}{{{N_T}{N_R}}})}$ and $ \prod\limits_{i,j}^{}{{F_Y}(\frac{ {{2^{\frac{2 R_s}{B}}} {x_2}}}{{{\Delta _1}{N_T}{N_R}}})}$ can be rewritten as
\begin{equation}
\prod\limits_{i,j}^{}{{F_Y}\left(\! \frac{ {{2^{\frac{2 R_s}{B}}} {x_1}}}{{{N_T}{N_R}}} \!\right)} = {\left(\! {\frac{{2^{\frac{2 R_s}{B}}}}{{ {N_T}{N_R}}}} \!\right)^{{N_T}{N_R}}}\!\prod\limits_{i,j}^{} {\frac{1}{{\sigma _{{s_{{m_i}}}{d_{{m_j}}}}^2}}} {x^{{N_T}{N_R}}}
\end{equation}
and
\begin{equation}
\prod\limits_{i,j}^{}{{F_Y}\!\!\left(\! \frac{ {{2^{\frac{2 R_s}{B}}} {x_1}}}{{{\Delta _1}{N_T}{N_R}}} \!\right)} = {\left(\! {\frac{{2^{\frac{2 R_s}{B}}}}{{ {\Delta _1}{N_T}{N_R}}}} \!\right)^{{N_T}{N_R}}}\!\!\!\!\!\!\prod\limits_{i,j}^{} {\frac{1}{{\sigma _{{s_{{m_i}}}{d_{{m_j}}}}^2}}} {x^{{N_T}{N_R}}},
\end{equation}
respectively.

Substituting (B.4) and (B.9) into (B.1) yields
\begin{eqnarray}
&&\!\!\!\!\!\!\!\!\!\!\!\!\!\!\!\!\!\!\!\!\!\!\!\!\!\!\!\!\!\Pr \left( {{{{N_T}{N_R}}}\mathop {\max }\limits_{i,j} {{\left| {{h_{{s_{{m_i}}}{d_{{m_j}}}}}} \right|}^2} < {2^{\frac{2 R_s}{B}}}\mathop {\max }\limits_{i,l} {{\left| {{h_{{s_{{m_i}}}{e_l}}}} \right|}^2}} \right) \nonumber \\
&&\!\!\!\!\!\!\!\!\!\!\!\!\!\!\!\!\!\!\!\!\!\!\!\!\!\!\!\!\!= \!\!\!\!\!\!\!\!\!\sum\limits_{n = 1}^{{2^{{N_T}{N_E}}} \!-\! 1} \!\!\!\!{{{\left(\!\! {\frac{{2^{\frac{2 R_s}{B}} }}{{ {N_T}{N_R}}}} \!\!\right)}^{{N_T}{N_R}}}\!\!\!\!\!\!\!\!\!\!\!\!\!\!\!\!\left(\! {{N_T}{N_R}} \!\right)!{{\left(\! { - 1} \!\right)}^{\left|\! {{C_n}} \!\right| \!+\! 1}}}\!\!{{{\left(\!\! {\sum\limits_{i,l \in {C_n}}^{} \!\!\!\!\!{\frac{1}{{\sigma _{{s_i}\!{e_l}}^2}}} } \!\!\right)}^{ - {N_T}{N_R}}}\!\!\!\!\!\!\!\!\!\!\!\!\!\!\!\!\!\!\!\!\!\!\prod\limits_{i,j}^{}\!\! {\frac{1}{{\sigma _{{s_{{m_i}}}\!{d_{{m_j}}}}^2}}} },
\end{eqnarray}
which can be further rewritten as
\begin{eqnarray}
&&\Pr \left( {{{{N_T}{N_R}}}\mathop {\max }\limits_{i,j} {{\left| {{h_{{s_{{m_i}}}{d_{{m_j}}}}}} \right|}^2} < {2^{\frac{2 R_s}{B}}}\mathop {\max }\limits_{i,l} {{\left| {{h_{{s_{{m_i}}}{e_l}}}} \right|}^2}} \right)\nonumber \\
&&=\!\! \sum\limits_{n = 1}^{{2^{{N_T}{N_E}}} \!-\! 1}{{{{\left(\! { - 1} \!\right)}^{\left| {{C_n}} \right|\! + \!1}}}{\omega _{il0}}{{\left(\!\! {\frac{1}{{{\lambda _{se}}}}} \!\!\right)}^{{N_T}{N_R}}}},
\end{eqnarray}
where $\omega _{il0} \!=\! (\!{N_T}{N_R}\!)!{{{(\! {\frac{{2^{\frac{2 R_s}{B}} }}{{ {N_T}\!{N_R}}}} \!)}^{{N_T}\!{N_R}}}}\!\!\!\!\!{{{{(\!\!\!\!\! {\sum\limits_{i,l \in {C_n}}^{} \!\!\!\!\!{\frac{1}{{\alpha _{{s_{m_i}}{e_l}}^{}}}} } \!)}^{-{N_T}\!{N_R}}}\!\!\!\!\!\!\!\!\!\!\!(\!\prod\limits_{i,j}^{} {\alpha _{{s_{{m_i}}}{d_{{m_j}}}}^{}}\!)^{-1} }}$.

Similarly to (B.12), (B.2) can be finally obtained as
\begin{eqnarray}
&&\!\!\!\!\!\!\!\!\!\!\!\!\!\!\Pr\! \left(\!\! \mathop {\max }\limits_{i,j} {{\left|\! {{h_{{s_{{m_i}}}{d_{{m_j}}}}}} \!\right|}^2} \!\!\!\!\!<\!\! \frac{{2^{\frac{2 R_s}{B}}}}{{{{N_T}\!{N_R}}}}{{\max}\!\!\left(\!\!\mathop {\max }\limits_{i,l} {{\left|\! {{h_{{s_{{m_i}}}{e_l}}}} \!\right|}^2}\!,\!\frac{1}{\Delta _1}\!\mathop {\max }\limits_{l}\! {{\left|\! {{h_{{s_{k}}{e_l}}}} \!\right|}^2}\!\!\right)} \!\!\right) \nonumber \\
&&\!\!\!\!\!\!\!\!\!\!\!\!\!\!=\!\!\! \sum\limits_{n = 1}^{{2^{{N_T}{N_E}}} \!-\! 1}\sum\limits_{g = 1}^{{2^{{N_E}}} \!-\! 1}\sum\limits_{t = 0}^{{N_T}{N_R}} \!\!\!{{{\left(\!-1\!\right)^{\left|\!C_n\!\right|\!+\!\left|\!F_g\!\right|}}}}{\alpha _{il0}}{{\left(\! {\frac{1}{{{\lambda _{se}}}}} \!\right)}^{{N_T}{N_R}}}\nonumber \\
&&\!\!\!\!\!\!\!\!\!+\!\!\!\sum\limits_{n = 1}^{{2^{{N_T}{N_E}}} \!-\! 1}\sum\limits_{g = 1}^{{2^{{N_E}}} \!-\! 1}\sum\limits_{t = 0}^{{N_T}{N_R}} \!\!\!{{{\left(\!-1\!\right)^{\left|\!C_n\!\right|\!+\!\left|\!F_g\!\right|}}}}{\beta _{il0}}{{\left(\! {\frac{1}{{{\lambda _{se}}}}} \!\right)}^{{N_T}{N_R}}}\!\!\!\!\!,
\end{eqnarray}
where ${\alpha _{il0}} = \frac{( {\sum\limits_{l \in {F_g}}^{} {\frac{1}{{\alpha _{{s_k}{e_l}}^{}}}} } ){(\!{N_T}{N_R}\!)!}\prod\limits_{i,j} {\frac{1}{{\alpha _{{s_{{m_i}}}{d_{{m_j}}}}^{}}}}{( {\frac{{{2^{{\frac{2 R_s}{B}}}}}}{{{N_T}{N_R}}}} )^{{N_T}{N_R}}}}{{\left(\!{N_T}{N_R}-t\!\right)!\left(\!\Delta _1\!\right)^t}{( {\sum\limits_{i,l \in {C_n}}^{} {\frac{1}{{\alpha _{{s_{{m_i}}}{e_l}}^{}}}} } )^{ {N_T}{N_R} - k}}{( {{\alpha _{il0}^{'}}} )^{ t + 1}}} $, ${\beta _{il0}} = \frac{( {\sum\limits_{i,l \in {C_n}}^{} {\frac{1}{{\alpha _{{s_{{m_i}}}{e_l}}^{}}}} } ){(\!{N_T}{N_R}\!)!}\prod\limits_{i,j} {\frac{1}{{\alpha _{{s_{{m_i}}}{d_{{m_j}}}}^{}}}}{( {\frac{{{2^{{\frac{2 R_s}{B}}}}}}{{{N_T}{N_R}{\Delta _1}}}} )^{{N_T}{N_R}}}}{{{\left(\!{N_T}{N_R}\!-\!t\!\right)!}{\left(\!\Delta _1\!\right)^{-t}}}{( {\sum\limits_{l \in {F_g}}^{} {\frac{1}{{\alpha _{{s_k}{e_l}}^{}}}} } )^{ {N_T}{N_R} - k}}{( {{\Delta _1}{\alpha _{il0}^{'}}} )^{ t + 1}}} $, and $\alpha _{il0}^{'}\!=\!\!\!\!\!{\sum\limits_{i,l \in {C_n}}^{} \!\!\!\!{\frac{1}{{\Delta _1} {\alpha _{{s_{{m_i}}}{e_l}}^{}}}}  +\!\!\!\! \sum\limits_{l \in {F_g}}^{}\!\!\! {\frac{1}{{\alpha _{{s_k}{e_l}}^{}}}} }$.

Based on (B.13) and (B.14), (B.1) can be reformulated as (B.15) shown at the top of the following page.
\begin{figure*}[t]
\begin{scriptsize}
\begin{equation}\label{equa46}
\begin{split}
 \begin{aligned}
 P_{\text{so}}^{\text{RSDPS}} \!\!\ge\!\! \frac{1}{M}\!\!\!\sum\limits_{m = 1}^M \!\!\!\left(\!\!\frac{1}{M}\!\!\left(\!\!{\sum\limits_{n = 1}^{{2^{{N_T}\!{N_E}}} \!-\! 1}\!\!\!\!\!\!{{{{\left(\! { - 1} \!\right)}^{\left|\! {{C_n}} \!\right|\! + \!1}}}{\omega _{il0}}}}{+\!\!\!\!\!\!\!\!\sum\limits_{n = 1}^{{2^{{N_T}\!{N_E}}} \!-\! 1}\!\sum\limits_{g = 1}^{{2^{{N_E}}} \!-\! 1}\!\sum\limits_{t = 0}^{{N_T}\!{N_R}} \!\!\!{{{\left(\!-1\!\right)^{\left|\!C_n\!\right|\!+\!\left|\!F_g\!\right|}}}}\bar P_{\text{o\_{km}}}{\alpha _{il0}}+\!\!\!\!\!\!\!\!\sum\limits_{n = 1}^{{2^{{N_T}\!{N_E}}} \!-\! 1}\!\sum\limits_{g = 1}^{{2^{{N_E}}} \!-\! 1}\!\sum\limits_{t = 0}^{{N_T}\!{N_R}} \!\!\!{{{\left(\!-1\!\right)^{\left|\!C_n\!\right|\!+\!\left|\!F_g\!\right|}}}}\bar P_{\text{o\_{km}}}{\beta _{il0}}}\!\!\right)\!\!\right)\!\!{{\left(\! {\frac{1}{{{\lambda _{se}}}}} \!\right)}^{{N_T}\!{N_R}}}\!\!\!.
\end{aligned}
\end{split}
\end{equation}
\end{scriptsize}
\hrule
\end{figure*}

Combining (34) and (B.15) yields
\begin{equation}
{d_{{\text{RSDPS}}}} \le {N_T}{N_R}.
\end{equation}

Furthermore, in the high-SNR region we can observe from the definition of $\Delta _0$ that as the transmit power $P_t$ tends to infinity, $\Delta _0$ approaches zero. Substituting the inequality $\sum\limits_{i = 1}^{{N_T}} {\sum\limits_{j = 1}^{{N_R}} {{{{{| {{h_{{s_{{m_i}}}{d_{{m_j}}}}}} |}^2}}} }  \ge \mathop {\max }\limits_{i,j} {| {{h_{{s_{{m_i}}}{d_{{m_j}}}}}} |^2}}$, ${2^{\frac{2 R_s}{B}}}\sum\limits_{i = 1}^{{N_T}} {\sum\limits_{l = 1}^{{N_E}} {{{{{| {{h_{{s_{{m_i}}}{e_l}}}} |}^2}}}} }+\Delta _0  \le {2^{\frac{2 R_s}{B}}}{{N_T}{N_E}}\mathop {\max }\limits_{i,l} {| {{h_{{s_{{m_i}}}{e_l}}}} |^2}$, and ${2^{\frac{2 R_s}{B}}}{\max}{(\sum\limits_{i = 1}^{{N_T}} {\sum\limits_{l = 1}^{{N_E}} {{{{{| {{h_{{s_{{m_i}}}{e_l}}}} |}^2}}}} },\frac{1}{\Delta _1} {\sum\limits_{l = 1}^{{N_E}} {{{{{| {{h_{{s_{{m_i}}}{e_l}}}} |}^2}}}} })}+\Delta _0  \le {2^{\frac{2 R_s}{B}}}{\max}{({{N_T}{N_E}}\mathop {\max }\limits_{i,l} {| {{h_{{s_{{m_i}}}{e_l}}}} |^2},\frac{{{N_E}}}{\Delta _1}\mathop {\max }\limits_{l} {| {{h_{{s_{k}}{e_l}}}} |^2})}$ into (18) and (19) yields
\begin{eqnarray}
&&\!\!\!\!\!\!\!\!\!\!\!\!\!\!\!\! P_{\text{so}}^{\text{RSDPS}} \!\!\le\!\! \frac{1}{M}\!\!\!\sum\limits_{m = 1}^M \!\!{\frac{1}{M}\!\!\left(\!\!\Pr\!\! {\left(\!\! {\mathop {\max }\limits_{i,j} \!{{\left|\! {{h_{{s_{{m_i}}}{d_{{m_j}}}}}} \!\right|}^2} \!\!\!\!<\!\! }\right.}\right.}{\left.{{2^{\frac{2 R_s}{B}}}{{{N_T}{N_E}}}\mathop {\max }\limits_{i,l}\! {{\left|\! {{h_{{s_{{m_i}}}{e_l}}}} \!\right|}^2}} \!\!\right)} \nonumber \\
&&+\!\!\!\!\!\!\!\!{\sum\limits_{k \in \mathbb{D} - \left\{ m \right\}}^{}\!\!\!\!\!\!\Pr\!\! \left(\!\! {\mathop {\max }\limits_{i,j} \!\!{{\left|\! {{h_{{s_{{m_i}}}{d_{{m_j}}}}}} \!\right|}^2} \!\!\!\!<\!\! }\right.}{{2^{\frac{2 R_s}{B}}}{\max}\!\!\left(\!\!{{{{N_T}{N_E}}}\mathop {\max }\limits_{i,l} {{\left|\! {{h_{{s_{{m_i}}}{e_l}}}} \!\right|}^2}},\right.}\nonumber \\
&&\;\;\;\;\;\;\;\;\;\;\;\;\left.{\left.{\left.\frac{{{{N_E}}}}{\Delta _1}{\mathop {\max }\limits_{l} {{\left|\! {{h_{{s_{k}}{e_l}}}} \!\right|}^2}}\!\!\right)} \!\!\right)}\!\!\right){\bar P_{\text {o\_{km}}}}.
\end{eqnarray}

Similarly to (B.15), (B.17) can be reformulated as (B.18) shown at the top of the following page, where $\omega _{il1} \!\!=\!\! (\!{N_T}\!{N_R}\!)!{{{(\! {{{2^{\frac{2 R_s}{B}} }}\!{{ {N_T}\!{N_E}}}} \!)}^{{N_T}\!{N_R}}}}\!\!{{{{(\!\!\! {\sum\limits_{i,l \in {C_n}}^{} \!\!\!\!{\frac{1}{{\alpha _{{s_{m_i}}{e_l}}^{}}}} } \!)}^{-{N_T}\!{N_R}}}\!\!\!\!\!(\!\prod\limits_{i,j}^{}\! {\alpha _{{s_{{m_i}}}{d_{{m_j}}}}^{}}\!)^{-1} }}$, ${\alpha _{il1}} \!=\!\! \frac{(\!\!\! {\sum\limits_{l \in {F_g}}^{} \!\!\!\!{\frac{1}{{\alpha _{{s_k}{e_l}}^{}}}} } \!)\prod\limits_{i,j}\! {\frac{1}{{\left(\!{N_T}{N_R}\!\right)!}{\alpha _{{s_{{m_i}}}{d_{{m_j}}}}^{}}}}{( {{{{2^{{\frac{2 R_s}{B}}}}}}{{{N_T}{N_E}}}} )^{{N_T}{N_R}}}}{{\left(\!{N_T}{N_R}-t\!\right)!\left(\!{N_T}{\Delta _1}\!\right)^t}{(\!\! {\sum\limits_{i,l \in {C_n}}^{} \!\!\!\!{\frac{1}{{\alpha _{{s_{{m_i}}}{e_l}}^{}}}} } \!)^{ {N_T}{N_R} \!-\! k}}{( {{\alpha _{il1}^{'}}} )^{ t \!+\! 1}}} $, $\alpha _{il1}^{'}\!=\!\!{\sum\limits_{i,l \in {C_n}}^{}\!\! {\frac{1}{{N_T}{\Delta _1} {\alpha _{{s_{{m_i}}}{e_l}}^{}}}}  \!+\!\!\! \sum\limits_{l \in {F_g}}^{} {\frac{1}{{\alpha _{{s_k}{e_l}}^{}}}} }$, and ${\beta _{il1}} \!\!=\!\! \frac{(\!\! {\sum\limits_{i,l \in {C_n}}^{} \!\!\!{\frac{1}{{\left(\!{N_T}{N_R}\!\right)!}{\alpha _{{s_{{m_i}}}{e_l}}^{}}}} } )\prod\limits_{i,j} {\frac{1}{{\alpha _{{s_{{m_i}}}{d_{{m_j}}}}^{}}}}{(\! {\frac{{{2^{{\frac{2 R_s}{B}}}}}{N_E}}{{{\Delta _1}}}} \!)^{{N_T}{N_R}}}}{{{\left(\!{N_T}{N_R}\!-\!t\!\right)!}{\left(\!{\Delta _1}{N_T}\!\right)^{-t}}}\!{(\! {\sum\limits_{l \in {F_g}}^{}\!\!\! {\frac{1}{{\alpha _{{s_k}{e_l}}^{}}}} } )^{ {N_T}{N_R} - k}}{( {{\Delta _1}{N_T}{\alpha _{il1}^{'}}} )^{ t + 1}}} $.
\begin{figure*}[t]
\begin{scriptsize}
\begin{equation}\label{equa49}
\begin{split}
 \begin{aligned}
 P_{\text{so}}^{\text{RSDPS}} \!\!\le\!\! \frac{1}{M}\!\!\!\sum\limits_{m = 1}^M \!\!\!\left(\!\!\frac{1}{M}\!\!\left(\!\!{\sum\limits_{n = 1}^{{2^{{N_T}\!{N_E}}} \!-\! 1}\!\!\!\!\!\!{{{{\left(\! { - 1} \!\right)}^{\left|\! {{C_n}} \!\right|\! + \!1}}}{\omega _{il1}}}}{+\!\!\!\!\!\!\!\sum\limits_{n = 1}^{{2^{{N_T}\!{N_E}}} \!-\! 1}\!\sum\limits_{g = 1}^{{2^{{N_E}}} \!-\! 1}\!\sum\limits_{t = 0}^{{N_T}\!{N_R}} \!\!\!{{{\left(\!-1\!\right)^{\left|\!C_n\!\right|\!+\!\left|\!F_g\!\right|}}}}\bar P_{\text{o\_{km}}}{\alpha _{il1}}+\!\!\!\!\!\!\!\!\sum\limits_{n = 1}^{{2^{{N_T}\!{N_E}}} \!-\! 1}\!\sum\limits_{g = 1}^{{2^{{N_E}}} \!-\! 1}\!\sum\limits_{t = 0}^{{N_T}\!{N_R}} \!\!\!{{{\left(\!-1\!\right)^{\left|\!C_n\!\right|\!+\!\left|\!F_g\!\right|}}}}\bar P_{\text{o\_{km}}}{\beta _{il1}}}\!\!\right)\!\!\right)\!\!{{\left(\! {\frac{1}{{{\lambda _{se}}}}} \!\right)}^{{N_T}\!{N_R}}}\!\!\!.
\end{aligned}
\end{split}
\end{equation}
\end{scriptsize}
\hrule
\end{figure*}

Moreover, substituting (B.18) into (34) yields
\begin{equation}
{d_{{\text{RSDPS}}}} \ge {N_T}{N_R}.
\end{equation}

Therefore, based on (B.16) and (B.19), the secrecy diversity gain of the conventional RSDPS scheme can be expressed as
\begin{equation}
{d_{{\text{RSDPS}}}} = {N_T}{N_R}.
\end{equation}

\emph{B, Proof of Theorem 2:}

Considering the inequality ${2^{\frac{2 R_s}{B}}}\!\! {\sum\limits_{l = 1}^{{N_E}}\!\! {{{{{| {{h_{{s_{{m_i}}}{e_l}}}} |}^2}}}} }\!+\!\Lambda _0  \!\!\ge\!\! {2^{\frac{2 R_s}{B}}}\!\mathop {\max }\limits_{l} {| {{h_{{s_{{m_i}}}{e_l}}}} |^2}$, \!\!\!\!$\mathop {\max }\limits_{{m \in \mathbb{D},1 \le i \le {N_T}}} \!\!{\sum\limits_{j = 1}^{{N_R}} \!\!{{{{{| {{h_{{s_{{m_i}}}{d_{{m_j}}}}}} \!|}^2}}} }  \!\!\le\!\! {{{N_R}}}\mathop {\max }\limits_{m,i,j} \!{| {{h_{{s_{{m_i}}}{d_{{m_j}}}}}} \!|^2}}$, and ${2^{\frac{2 R_s}{B}}}{\max}{(\!\! {\sum\limits_{l = 1}^{{N_E}}\!\! {{{{{| {{h_{{s_{{m_i}}}{e_l}}}} |}^2}}}} }\!,\!\frac{1}{\Lambda _1} \!\!{\sum\limits_{l = 1}^{{N_E}} \!\!{{{{{| {{h_{{s_{{m_i}}}{e_l}}}} |}^2}}}} })}+\Lambda _0  \ge {2^{\frac{2 R_s}{B}}}{\max}{(\mathop {\max }\limits_{l} {| {{h_{{s_{{m_i}}}{e_l}}}} |^2},\frac{1}{\Lambda _1}\mathop {\max }\limits_{l} {| {{h_{{s_{k}}{e_l}}}} |^2})}$, we arrive at
\begin{eqnarray}
&&\!\!\!\!\!\!\!\!\!\!\!\! P_{\text{so}}^{\text{TAS}} \!\!\ge\!\! {\frac{1}{M}\left(\!\!\Pr \!\!{\left(\!\! {{{{N_R}}}\mathop {\max }\limits_{m,i,j} {{\left| {{h_{{s_{{m_i}}}{d_{{m_j}}}}}} \right|}^2} \!\!< }\right.}\right.}{\left.{{2^{\frac{2 R_s}{B}}}\mathop {\max }\limits_{l} {{\left| {{h_{{s_{{m_i}}}{e_l}}}} \right|}^2}} \right)} \nonumber \\
&&+\!\!\!\!\!\!\!{\sum\limits_{k \in \mathbb{D} - \left\{ m \right\}}^{}\!\!\!\!\!\!\Pr\!\! \left(\!\! {{{{N_R}}}\mathop {\max }\limits_{m,i,j} {{\left| {{h_{{s_{{m_i}}}{d_{{m_j}}}}}} \right|}^2} \!\!\!\!<\!\! }\right.}{{2^{\frac{2 R_s}{B}}}{\max}\left({\mathop {\max }\limits_{l} {{\left| {{h_{{s_{{m_i}}}{e_l}}}} \right|}^2}}\right.},\nonumber \\
&&\;\;\;\;\;\;\;\;\;\;\;\;\;\left.{\left.{\left.\frac{1}{\Lambda _1}{\mathop {\max }\limits_{l} {{\left| {{h_{{s_{k}}{e_l}}}} \right|}^2}}\!\!\right)} \!\!\right)}\!\!\right){\bar P_{\text{o\_{km}}}}.
\end{eqnarray}

Similarly to (B.12) and (B.14), $\Pr ( {{{{N_T}{N_R}}}\mathop {\max }\limits_{m,i,j} {{| {{h_{{s_{{m_i}}}{d_{{m_j}}}}}} |}^2}} \\{< {2^{\frac{2 R_s}{B}}}\mathop {\max }\limits_{i,l} {{| {{h_{{s_{{m_i}}}{e_l}}}} |}^2}} )$ and $\Pr ( {{{{N_T}{N_R}}}\mathop {\max }\limits_{m,i,j} {{| {{h_{{s_{{m_i}}}{d_{{m_j}}}}}} |}^2} <{2^{\frac{2 R_s}{B}}} }\\{{{\max}(\mathop {\max }\limits_{i,l} {{| {{h_{{s_{{m_i}}}{e_l}}}} |}^2},}}{{\frac{1}{\Delta _1}\mathop {\max }\limits_{l} {{| {{h_{{s_{k}}{e_l}}}} |}^2})}} )$ can be rewritten as
\begin{eqnarray}
&&\Pr \left(\!\! {\mathop {\max }\limits_{m,i,j} {{\left| {{h_{{s_{{m_i}}}{d_{{m_j}}}}}} \right|}^2}} \!\!<\!\! \frac{1}{{{{N_T}{N_R}}}}\left(\!\!{{2^{\frac{2 R_s}{B}}}\mathop {\max }\limits_{i,l} {{\left| {{h_{{s_{{m_i}}}{e_l}}}} \right|}^2}}\!\!\right)\!\! \right)\nonumber\\
&&= \sum\limits_{n = 1}^{{2^{{N_T}\!{N_E}}} \!- \!1}{{{{{\left(\! { - 1} \!\right)}^{\left| {{C_n}} \right| \!+\! 1}}}}\omega _{mil0}{{\left(\!\! {\frac{1}{{{\lambda _{se}}}}} \!\!\right)}^{M\!{N_T}\!{N_R}}}}
\end{eqnarray}
and
\begin{eqnarray}
&&\!\!\!\!\!\!\!\!\!\!\!\!\!\Pr \left( {\mathop {\max }\limits_{m,i,j} {{\left| {{h_{{s_{{m_i}}}{d_{{m_j}}}}}} \right|}^2} < \frac{1}{{{{N_T}{N_R}}}}{2^{\frac{2 R_s}{B}}}}\right.\nonumber \\
&&\left.{{\max}\left(\mathop {\max }\limits_{i,l} {{\left| {{h_{{s_{{m_i}}}{e_l}}}} \right|}^2},\frac{1}{\Delta _1}\mathop {\max }\limits_{l} {{\left| {{h_{{s_{k}}{e_l}}}} \right|}^2}\right)} \right) \nonumber \\
&&\!\!\!\!\!\!\!\!\!\!\!\!\!=\!\!\! \sum\limits_{n = 1}^{{2^{{N_T}\!{N_E}}} \!-\! 1}\sum\limits_{g = 1}^{{2^{{N_E}}} \!-\! 1}\sum\limits_{t = 0}^{M\!{N_T}\!{N_R}} \!\!\!{{{\left(\!-1\!\right)^{\left|\!C_n\!\right|\!+\!\left|\!F_g\!\right|}}}}{\alpha _{mil0}}{{\left(\!\! {\frac{1}{{{\lambda _{se}}}}} \!\!\right)}^{M\!{N_T}\!{N_R}}} \nonumber \\ &&\!\!\!\!\!\!\!+\!\!\!\!\!\!\!\sum\limits_{n = 1}^{{2^{{N_T}\!{N_E}}} \!-\! 1}\sum\limits_{g = 1}^{{2^{{N_E}}} \!-\! 1}\sum\limits_{t = 0}^{M\!{N_T}\!{N_R}} \!\!\!{{{\left(\!-1\!\right)^{\left|\!C_n\!\right|\!+\!\left|\!F_g\!\right|}}}}{\beta _{mil0}}{{\left(\!\! {\frac{1}{{{\lambda _{se}}}}} \!\!\right)}^{M\!{N_T}\!{N_R}}}\!\!\!\!\!\!\!\!\!\!\!\!\!\!,
\end{eqnarray}
respectively, where ${\alpha _{mil0}} \!\!=\!\! \frac{{(\!M\!{N_T}\!{N_R}\!)!}(\! {\sum\limits_{l \in {F_g}}^{}\!\!\! {\frac{1}{{\alpha _{{s_k}{e_l}}^{}}}} } \!)\!\!\!\prod\limits_{m,i,j}\!\!\!\! {\frac{1}{{\alpha _{{s_{{m_i}}}{d_{{m_j}}}}^{}}}}{( {\frac{{{2^{{\frac{2 R_s}{B}}}}}}{{{N_T}{N_R}}}} )^{M{N_T}{N_R}}}}{{(\!M\!{N_T}\!{N_R}\!-\!t\!)!(\!\Delta _1\!)^t}{(\!\! {\sum\limits_{i,l \in {C_n}}^{} \!\!\!\!\!{\frac{1}{{\alpha _{{s_{{m_i}}}{e_l}}^{}}}} } )^{ M{N_T}{N_R} - t}}{( {{\alpha _{il0}^{'}}} )^{ t + 1}}} $, $\omega _{mil0} \!=\! (\!M{N_T}{N_R}\!)!{{{(\! {\frac{{2^{\frac{2 R_s}{B}} }}{{ {N_T}\!{N_R}}}} \!)}^{M\!{N_T}\!{N_R}}}}{{{{(\!\! {\sum\limits_{i,l \in {C_n}}^{} \!\!{\frac{1}{{\alpha _{{s_i}{e_l}}^{}}}} } \!)}^{-M\!{N_T}\!{N_R}}}\!\!(\!\prod\limits_{m,i,j}^{} {\alpha _{{s_{{m_i}}}{d_{{m_j}}}}^{}}\!)^{-1} }}$, and ${\beta _{mil0}} \!=\! \frac{{(\!M\!{N_T}\!{N_R}\!)!}(\!\!\! {\sum\limits_{i,l \in {C_n}}^{} \!\!\!{\frac{1}{{\alpha _{{s_{{m_i}}}{e_l}}^{}}}} } \!)\!\!\prod\limits_{m,i,j}\!\!\!\! {\frac{1}{{\alpha _{{s_{{m_i}}}{d_{{m_j}}}}^{}}}}{( {\frac{{{2^{{\frac{2 R_s}{B}}}}}}{{{N_T}{N_R}{\Delta _1}}}} )^{M{N_T}{N_R}}}}{{{(\!M\!{N_T}\!{N_R}\!-\!t\!)!}{(\!\Delta _1\!)^{-t}}}{(\!\! {\sum\limits_{l \in {F_g}}^{}\!\!\! {\frac{1}{{\alpha _{{s_k}{e_l}}^{}}}} } \!\!)^{ M{N_T}{N_R} \!-\! t}}{(\! {{\Delta _1}{\alpha _{il0}^{'}}} \!)^{ t + 1}}} $.

With the aid of (B.22) and (B.23), we arrive at (B.24) shown at the top of the following page, where $\omega _{mil2} \!=\! (\!M\!{N_T}\!{N_R}\!)!{{{(\! {\frac{{2^{\frac{2 R_s}{B}} }}{{ {N_R}}}} \!)}^{M\!{N_T}\!{N_R}}}}\!\!\!\!{{{{(\!\!\!\! {\sum\limits_{i,l \in {C_n}}^{} \!\!\!\!{\frac{1}{{\alpha _{{s_i}{e_l}}^{}}}} } \!)}^{-M\!{N_T}\!{N_R}}}\!\!(\!\prod\limits_{i,j}^{} \!{\alpha _{{s_{{m_i}}}{d_{{m_j}}}}^{}}\!)^{-1} }}$, ${\alpha _{mil2}} \!=\! \frac{{(\!M\!{N_T}\!{N_R}\!)!}(\!\! {\sum\limits_{l \in {F_g}}^{}\!\!\!\! {\frac{1}{{\alpha _{{s_k}{e_l}}^{}}}} } \!)\!\prod\limits_{m,i,j}\!\! {\frac{1}{{\alpha _{{s_{{m_i}}}{d_{{m_j}}}}^{}}}}{( {\frac{{{2^{{\frac{2 R_s}{B}}}}}}{{{N_R}}}} )^{M{N_T}{N_R}}}}{{(\!M\!{N_T}\!{N_R}\!-\!t\!)!(\!\Lambda _1\!)^t}{(\!\! {\sum\limits_{l \in {C_n}}^{} \!\!\!{\frac{1}{{\alpha _{{s_{{m_i}}}{e_l}}^{}}}} } )^{ M{N_T}{N_R} - t}}{( {{\alpha _{il2}^{'}}} )^{ t + 1}}} $, $\alpha _{il2}^{'}\!\!=\!\!\!\!\!{\sum\limits_{l \in {C_n}}^{}\!\!\! {\frac{1}{{\Lambda _1} {\alpha _{{s_{{m_i}}}{e_l}}^{}}}} \!\! +\!\!\!\!\! \sum\limits_{l \in {F_g}}^{} \!\!\!{\frac{1}{{\alpha _{{s_k}{e_l}}^{}}}} }$, and ${\beta _{mil2}} \!=\! \frac{{(\!M\!{N_T}\!{N_R}\!)!}(\! {\sum\limits_{l \in {C_n}}^{} \!\!\!{\frac{1}{{\alpha _{{s_{{m_i}}}{e_l}}^{}}}} } )\!\!\prod\limits_{m,i,j}\!\! {\frac{1}{{\alpha _{{s_{{m_i}}}{d_{{m_j}}}}^{}}}}{( {\frac{{{2^{{\frac{2 R_s}{B}}}}}}{{{N_R}{\Lambda _1}}}} )^{M{N_T}{N_R}}}}{{{(\!M\!{N_T}\!{N_R}\!-\!t\!)!}{(\!\Lambda _1\!)^{-t}}}{( {\sum\limits_{l \in {F_g}}^{} {\frac{1}{{\alpha _{{s_k}{e_l}}^{}}}} } )^{ M{N_T}{N_R} - t}}{( {{\Lambda _1}{\alpha _{il2}^{'}}} )^{ t + 1}}} $.
\begin{figure*}[t]
\begin{scriptsize}
\begin{equation}\label{equa62}
\begin{split}
 \begin{aligned}
 P_{\text{so}}^{\text{TAS}} \!\!\ge\!\!  \frac{1}{M}\!\!\left(\!\!{\sum\limits_{n = 1}^{{2^{{N_E}}} \!-\! 1}\!\!\!{{{{\left(\! { - 1} \!\right)}^{\left| {{C_n}} \right|\! + \!1}}}{\omega _{mil2}}}}{+\!\!\!\!\sum\limits_{n = 1}^{{2^{{N_E}}} \!-\! 1}\!\sum\limits_{g = 1}^{{2^{{N_E}}} \!-\! 1}\!\sum\limits_{t = 0}^{M{N_T}{N_R}} \!\!\!\!\!\!\!{{{\left(\!-1\!\right)^{\left|\!C_n\!\right|\!+\!\left|\!F_g\!\right|}}}}\bar P_{\text{o\_{km}}}{\alpha _{mil2}}+\!\!\!\!\sum\limits_{n = 1}^{{2^{{N_E}}} \!-\! 1}\!\sum\limits_{g = 1}^{{2^{{N_E}}} \!-\! 1}\!\sum\limits_{t = 0}^{M{N_T}{N_R}} \!\!\!\!\!\!\!{{{\left(\!-1\!\right)^{\left|\!C_n\!\right|\!+\!\left|\!F_g\!\right|}}}}\bar P_{\text{o\_{km}}}{\beta _{mil2}}}\!\!\right)\!\!{{\left(\! {\frac{1}{{{\lambda _{se}}}}} \!\right)}^{M{N_T}{N_R}}}.
\end{aligned}
\end{split}
\end{equation}
\end{scriptsize}
\hrule
\end{figure*}
Substituting (B.24) into (36) yields
\begin{equation}
{d_{\text{TAS}}} \le M{N_T}{N_R}.
\end{equation}

Furthermore, upon considering an infinite SNR and using the inequality $\mathop {\max }\limits_{{m \in \mathbb{D},1 \le i \le {N_T}}} \!\!{\sum\limits_{j = 1}^{{N_R}}\!\! {{{{{| {{h_{{s_{{m_i}}}{d_{{m_j}}}}}} \!|}^2}}} }  \!\!\ge}{ \mathop {\max }\limits_{m,i,j}\! {| {{h_{{s_{{m_i}}}{d_{{m_j}}}}}} |^2}}$, ${2^{\frac{2 R_s}{B}}} {\sum\limits_{l = 1}^{{N_E}} {{{{{| {{h_{{s_{{m_i}}}{e_l}}}} |}^2}}}} }+\Lambda _0  \le {2^{\frac{2 R_s}{B}}}{{N_E}}\mathop {\max }\limits_{l} {| {{h_{{s_{{m_i}}}{e_l}}}} |^2}$, and ${2^{\frac{2 R_s}{B}}}{\max}{( {\sum\limits_{l = 1}^{{N_E}} {{{{{| {{h_{{s_{{m_i}}}{e_l}}}} |}^2}}}} },\frac{1}{\Lambda _1} {\sum\limits_{l = 1}^{{N_E}} {{{{{| {{h_{{s_{{m_i}}}{e_l}}}} |}^2}}}} })}+\Lambda _0  \le {2^{\frac{2 R_s}{B}}}$\\${\max}{({{N_E}}\mathop {\max }\limits_{l} {| {{h_{{s_{{m_i}}}{e_l}}}} |^2},\frac{{{N_E}}}{\Lambda _1}\mathop {\max }\limits_{l} {| {{h_{{s_{k}}{e_l}}}} |^2})}$, we have
\begin{eqnarray}
&&\!\!\!\!\!\!\!\!\!\!\!\!\! P_{\text{so}}^{\text{TAS}}\! \le \! {\frac{1}{M}\left(\!\Pr {\left(\! {\mathop {\max }\limits_{m,i,j} {{\left| {{h_{{s_{{m_i}}}{d_{{m_j}}}}}} \right|}^2} \!<\! {{2^{\frac{2 R_s}{B}}}{N_E}}}\right.}\right.}{\left.{\mathop {\max }\limits_{l} {{\left| {{h_{{s_{{m_i}}}{e_l}}}} \right|}^2}} \!\!\right)}\nonumber \\
&&+\!\!\!\!\!\!\!{\sum\limits_{k \in \mathbb{D} - \left\{ m \right\}}^{}\!\!\!\!\!\!\!\Pr\! \left(\!\! {\mathop {\max }\limits_{m,i,j} {{\left| {{h_{{s_{{m_i}}}{d_{{m_j}}}}}} \right|}^2} \!\!\!\!<\!\! {{2^{\frac{2 R_s}{B}}}{N_E}}}\right.}{{\max}\left({\mathop {\max }\limits_{l} {{\left| {{h_{{s_{{m_i}}}{e_l}}}} \right|}^2}},\right.}\nonumber \\
&&\;\;\;\;\;\;\;\;\;\;\;\;\left.{\left.{\left.\frac{1}{\Lambda _1}{\mathop {\max }\limits_{l} {{\left| {{h_{{s_{k}}{e_l}}}} \right|}^2}}\right)} \right)}\right){\bar P_{\text{{o\_{km}}}}}.
\end{eqnarray}

Similarly to (B.21), (B.26) can be expanded as (B.27) shown at the top of the following page, where ${\alpha _{mil3}} \!=\! \frac{{(\!M\!{N_T}\!{N_R}\!)!}(\! {\sum\limits_{l \in {F_g}}^{} \!\!\!{\frac{1}{{\alpha _{{s_k}{e_l}}^{}}}} }\! )\!\!\prod\limits_{m,i,j} \!\!\!\!\!{\frac{1}{{\alpha _{{s_{{m_i}}}{d_{{m_j}}}}^{}}}}{( {{{{2^{{\frac{2 R_s}{B}}}}}}{{{N_E}}}} )^{M{N_T}{N_R}}}}{{(\!M\!{N_T}\!{N_R}\!-\!t\!)!(\!{\Lambda _1}\!)^t}{(\!\! {\sum\limits_{l \in {C_n}}^{}\!\!\! {\frac{1}{{\alpha _{{s_{{m_i}}}{e_l}}^{}}}} } \!)^{ M{N_T}{N_R} - t}}{( {{\alpha _{il2}^{'}}} )^{ t + 1}}} $, $\omega _{mil3} \!=\! (\!M{N_T}{N_R}\!)!{{{(\! {{{2^{\frac{2 R_s}{B}} }}{{ {N_E}}}} \!)}^{M\!{N_T}\!{N_R}}}}\!\!{{{{(\!\! {\sum\limits_{l \in {C_n}}^{} \!\!\!\!\!{\frac{1}{{\alpha _{{s_{m_i}}{e_l}}^{}}}} } \!)}^{-M\!{N_T}\!{N_R}}}}}\!\!\!{{(\!\!\!\prod\limits_{m,i,j}^{} \!\!\!{\alpha _{{s_{{m_i}}}{d_{{m_j}}}}^{}}\!)^{-1} }}$ and ${\beta _{mil3}} = \frac{{(\!M\!{N_T}\!{N_R}\!)!}( {\sum\limits_{l \in {C_n}}^{} {\frac{1}{{\alpha _{{s_{{m_i}}}{e_l}}^{}}}} } )\!\!\prod\limits_{m,i,j} \!\!{\frac{1}{{\alpha _{{s_{{m_i}}}{d_{{m_j}}}}^{}}}}{( {\frac{{{2^{{\frac{2 R_s}{B}}}}{N_E}}}{{{\Lambda _1}}}} )^{M{N_T}{N_R}}}}{{{(\!M\!{N_T}\!{N_R}\!-\!t\!)!}{(\!{\Lambda _1}\!)^{-t}}}{( {\sum\limits_{l \in {F_g}}^{} {\frac{1}{{\alpha _{{s_k}{e_l}}^{}}}} } )^{ M{N_T}{N_R} - t}}{( {{\Lambda _1}{\alpha _{il2}^{'}}} )^{ t + 1}}} $.
\begin{figure*}[t]
\begin{scriptsize}
\begin{equation}\label{equa65}
\begin{split}
 \begin{aligned}
 P_{\text{so}}^{\text{TAS}} \!\!\le\!\!  \frac{1}{M}\!\!\left(\!\!{\sum\limits_{n = 1}^{{2^{{N_E}}} \!-\! 1}\!\!\!{{{{\left(\! { - 1} \!\right)}^{\left| {{C_n}} \right|\! + \!1}}}{\omega _{mil3}}}}{+\!\!\!\!\sum\limits_{n = 1}^{{2^{{N_E}}} \!-\! 1}\!\sum\limits_{g = 1}^{{2^{{N_E}}} \!-\! 1}\!\sum\limits_{t = 0}^{M{N_T}{N_R}} \!\!\!\!\!\!\!{{{\left(\!-1\!\right)^{\left|\!C_n\!\right|\!+\!\left|\!F_g\!\right|}}}}\bar P_{\text{o\_{km}}}{\alpha _{mil3}}+\!\!\!\!\sum\limits_{n = 1}^{{2^{{N_E}}} \!-\! 1}\!\sum\limits_{g = 1}^{{2^{{N_E}}} \!-\! 1}\!\sum\limits_{t = 0}^{M{N_T}{N_R}} \!\!\!\!\!\!\!{{{\left(\!-1\!\right)^{\left|\!C_n\!\right|\!+\!\left|\!F_g\!\right|}}}}\bar P_{\text{o\_{km}}}{\beta _{mil3}}}\!\!\right)\!\!{{\left(\! {\frac{1}{{{\lambda _{se}}}}} \!\right)}^{M{N_T}{N_R}}}.
\end{aligned}
\end{split}
\end{equation}
\end{scriptsize}
\hrule
\end{figure*}
Hence, upon using (36) and (B.27), we obtain

\begin{equation}
{d_{\text{TAS}}} \ge M{N_T}{N_R}.
\end{equation}

By combining (B.25) and (B.28), we arrive at the secrecy diversity gain of the proposed TAS-SDPS scheme as
\begin{equation}
{d_{\text{TAS}}} = M{N_T}{N_R}.
\end{equation}

\ifCLASSOPTIONcaptionsoff
  \newpage
\fi

\end{spacing}
\end{document}